\documentclass{sig-alternate}
\pdfoutput=1

\usepackage{multirow}
\usepackage{algorithm}
\usepackage{algorithmic}
\usepackage{subfigure}
\usepackage{balance}
\usepackage{amsmath}
\usepackage{xcolor}
\usepackage{amsfonts}
\usepackage{amssymb}
\usepackage{graphicx}

\begin{document}
%
\conferenceinfo{WOODSTOCK}{'97 El Paso, Texas USA}

\title{Structural Learning of Diverse Ranking}

%
%
%
%
%

\numberofauthors{1} 
%
\author{
%
%
\alignauthor {Yadong Zhu\hspace*{0.5cm}Yanyan Lan\hspace*{0.5cm}Jiafeng Guo}\hspace*{0.5cm}Xueqi Cheng\\
\affaddr{Institute of Computing Technology, Chinese Academy of Sciences, Beijing 100190, China} \\
\email{\{zhuyadong\}@software.ict.ac.cn} \\
\email{\{lanyanyan, guojiafeng, cxq\}@ict.ac.cn }
}


\maketitle
\begin{abstract}
Relevance and diversity are both crucial criteria for an effective search system. In this paper, we propose a unified learning framework for simultaneously optimizing both relevance and diversity. Specifically, the problem is formalized as a structural learning framework optimizing Diversity-Correlated Evaluation Measures (DCEM), such as ERR-IA, $\alpha$-NDCG and NRBP. Within this framework, the discriminant function is defined to be a bi-criteria objective maximizing the sum of the relevance scores and dissimilarities (or diversity) among the documents. Relevance and diversity features are utilized to define the relevance scores and dissimilarities, respectively. Compared with traditional methods, the advantages of our approach lie in that: (1) Directly optimizing DCEM as the loss function is more fundamental for the task; (2) Our framework does not rely on explicit diversity information such as subtopics, thus is more adaptive to real application; (3) The representation of diversity as the feature-based scoring function is more flexible to incorporate rich diversity-based features into the learning framework. Extensive experiments on the public TREC datasets show that our approach significantly outperforms state-of-the-art diversification approaches, which validate the above advantages.
\end{abstract}

\category{H.3.3}{Information Search and Retrieval}{Information
Search and Retrieval -- \textit{Retrieval Models}}

\terms{Algorithms, Experimentation, Performance, Theory}

\keywords{Discriminant Function, Diversity Feature, Learning Framework}

\section{Introduction}

Relevance and diversity are both critical for user experiences in the real search scenario. On one hand, more relevant items should be ranked higher to satisfy users' information need. On the other hand, redundant information should be reduced to satisfy users' diverse information need. Recently, many diversity-correlated IR evaluation measures have been proposed for evaluating the search systems from the two aspects,
such as ERR-IA \cite{Agrawal:2009,ERR}, $\alpha$-NDCG \cite{Clarke:2008} and NRBP \cite{NRBP}, which try to achieve a balance between  relevance and diversity.

To fulfill the requirements of both relevance and diversity, many diversity-enhancement methods have been developed, and can be mainly divided into two categories: implicit and explicit methods \cite{Santos:ESR}.
Implicit methods such as MMR \cite{MMR} conduct greedy process based on heuristic defined objectives to select documents.
While explicit methods such as the work in \cite{Agrawal:2009,Santos:2010},
directly diversify search results based on the subtopic information of user queries, and then greedily select documents according to their predefined utility functions.

However, there are some disadvantages with these approaches: (1) the objectives of implicit methods are mainly heuristic, and it is not clear about the relations between them and the DCEM measures; (2) the diversity in explicit methods is achieved through the representation of subtopic information, thus is very easy to introduce bias in the estimation process of subtopics; (3) these approaches often utilize a predefined utility function, and thus limited features can be incorporated for capturing relevance and diversity properly.

In order to tackle the above challenges, we propose a unified learning framework to simultaneously optimize relevance and diversity in this paper. Firstly, we formalize the problem as a structural learning framework, in which the objective functions are directly defined as the diversity-correlated IR evaluation measures, such as ERR-IA, $\alpha$-NDCG and NRBP.
Secondly, we define the discriminant function as a bi-criteria objective, which maximizes the sum of the relevance scores and dissimilarities among the documents. Thirdly, we propose a bunch of features to capture relevance and diversity.
For relevance, the traditional relevance features used in learning-to-rank literature \cite{qin:letor} are adopted;
for diversity, a series of diversity features are utilized, such as dissimilarities of implicit topics, titles, texts, links, urls.

To evaluate the effectiveness of the proposed approach, we conduct extensive experiments on the public TREC datasets. The experimental results show that our methods can significantly outperform the state-of-the-art diversification approaches with the evaluation of ERR-IA, $\alpha$-NDCG and NRBP. Furthermore, our methods also achieve best in the evaluations of traditional intent-aware measures, i.e. Precision-IA and Subtopic recall. In addition, we give some discussions on the robustness of our methods and the importance of the proposed diversity features. Finally, we also study the efficiency of our approach based on the analysis of running time.

The main contributions of this paper lie in:
\begin{itemize}
\item  the proposal of a unified learning framework to simultaneously optimize both relevance and diversity.
\item  the definition of the discriminant function as a bi-criteria objective.
\item  the proposal of rich useful diversity-based features.
\item  a thorough experimental evaluation of the proposed approach and numerous baseline methods.
\end{itemize}

The rest of the paper is organized as follows. Section 2 describes the related work on search result diversification.
Section 3 formulates the learning problem of diversity-combined rankings.
Section 4 describes the formulation of discriminant function based on a bi-criteria objective, and presents a series of useful diversity features.
Section 5 describes the training procedure based on the structural SVM framework.
Section 6 contains the experimental setup and results. Section 7 presents our concluding remarks.

\section{Related Work}

In this section, we review the research work on search diversification. In general, they can be divided into two categories:
diversity-correlated methods and diversity-correlated evaluation measures. We will introduce more detailed information in the following.

\subsection{Diversity-Correlated Methods}

Diversity-correlated methods can be mainly divided into two categories: implicit approaches and explicit approaches \cite{Santos:ESR}.
The implicit methods assume that similar documents cover similar aspects and model inter-document dependencies. For example, Maximal Marginal Relevance (MMR) method \cite{MMR} proposes to iteratively select a candidate document with the highest similarity to the user query and the lowest similarity to the already selected documents, in order to promote novelty. In fact, most of the existing approaches are somehow inspired by the MMR method.
Zhai et al. \cite{Zhai:2003} select documents with high divergence from one language model to another based on the risk minimization consideration.
The explicit methods explicitly model aspects of a query and then select documents that cover different aspects. The aspects of a user query can be achieved with a taxonomy \cite{Agrawal:2009,Intent-aware,Explicit}, top retrieved documents \cite{Carterette:2009:PMR}, query reformulations \cite{Radlinski:2006,Santos:2010}, or multiple external resources \cite{He:2012}.
Overall, the explicit methods have shown better experimental performances comparing with implicit methods.

There are also some other methods which attempt to borrow theories from economical or political domains \cite{Portfolio,Rafiei:2010,proportionality}. The work in \cite{Portfolio,Rafiei:2010} applies economical portfolio theory for search result ranking, which views search diversification as a means of risk minimization. The approach in \cite{proportionality} treats the problem of finding a diverse search result as finding a proportional representation for the document ranking, which is like a critical part of most electoral processes.

Recently, some researchers have proposed to utilize machine learning techniques to solve the diversification problem.
Yue et al. \cite{Yue:2008} propose to optimize subtopic coverage as the loss function, and formulate a discriminant function based on maximizing word coverage. However, their work only focuses on diversity, and discards the requirements of relevance.
They claim that modeling both relevance and diversity simultaneously is a more challenging problem, which is exactly what we try to tackle in this paper.
Yadong et al. \cite{zhu:2014:rRLTR} propose a R-LTR model to solve the diversification problem, which comes from a sequential ranking process. Our work in this paper try to solve the diversification problem from a discriminative view, which is a unified framework.
The authors of \cite{Brandt:2011:DRR,two-level} try to construct a dynamic ranked-retrieval model, which may be useful in the user interface designing of future retrieval system. Our paper focuses on the common static ranking scenario, which is different from their papers.

There are also some on-line learning methods that try to learn retrieval models by exploiting user click data or implicit user feedback \cite{Radlinski:2008:LDR,icml/SlivkinsRG10,Raman:2012:OLD2}. These research work can tackle diversity problem to some extent, but they focus on an `on-line' or `interactive' scenario, which is different from our work. For example, Raman et al. \cite{Raman:2012:OLD2} propose an online algorithm that presents a ranking to users at each step, and observes the set of documents the user reads in the presented ranking, and then updates its model. While in our work, we try to utilize human labels conveying relevance and subtopic information to learn an optimal `off-line' retrieval model. The most representative scenario is the \textit{diversity task} of Web Track in TREC. In fact, the two modes (i.e.~`off-line' and `on-line') are complementary in practical applications. People usually utilize historical human labeled data to train an optimal retrieval model, and then use on-line user feedback to update the retrieval model dynamically. We may investigate on-line algorithms in our future work.

In this paper, we propose to utilize machine learning techniques to simultaneously optimizing both relevance and diversity based on a bi-criteria objective, which is different from traditional learning-based approaches and shows promising experimental performance.

\subsection{Diversity-Correlated Evaluation Measures}
The proper evaluation measures are very important to the diversity problem. They usually act as objective functions to be optimized by retrieval systems. However, traditional evaluation measures cannot well capture the diversity property. Therefore, several research studies on diversity evaluation measure have been proposed.

In the early stage, Zhai et al. \cite{Zhai:2003} define a number of subtopic recall metrics to measure diversity. Recently, many evaluation measures based on cascade models have been proposed, such as $\alpha$-NDCG \cite{Clarke:2008}, ERR-IA \cite{ERR}, and NRBP \cite{NRBP}. They measure the diversity of a result list by explicitly rewarding novelty and penalizing redundancy observed at every rank.
In the meantime, Agrawal et al. \cite{Agrawal:2009} also propose a series of \textit{intent-aware} versions of the traditional measures, such as MAP-IA, Precision-IA. The traditional measures are applied to each subtopic independently and then combined together.
More recently, Sakai and Song compare a wide range of diversified IR metrics, and propose a series of $D\#$ measures which have high discriminative power \cite{Sakai:2011,Sakai:2012:EIN}.
Interestingly, a novel proportionality measure called CPR (Cumulative Proportionality measure) has been proposed \cite{proportionality}, which captures proportionality in search results.

Overall, how to evaluate diversity properly is still an interesting research problem. The current official evaluation measures of TREC diversity task are ERR-IA, $\alpha$-NDCG and NRBP \cite{TREC11}, which are also the main objectives to be optimized in our work.
As summarized in \cite{comparative}, they are all based on cascade models and have the same nature. Moreover, the ERR-IA measure enables graded relevance values more than binary relevance.

\section{The Learning Problem}

\begin{table*}[t]
\centering
\caption{Summary of typical DCEM measures \cite{comparative}}
\begin{tabular}{c|c|c|c|c} \hline
diversity&novelty&gain&discount&measure\\ \hline
\multicolumn{1}{c|}{\multirow{3}{*}{$\mathcal{S}=\frac{\sum_{i=1}^{M}p_i \mathcal{S}_i}{\mathcal{N}}$}}  & \multicolumn{1}{c|}{\multirow{3}{*}{$\mathcal{S}_i=\sum_{k=1}^{K}\frac{Q_i^k}{D_k}$}} &
\multicolumn{1}{c|}{\multirow{1}{*}{$Q_i^k = q_i^k \prod_{j=1}^{k-1}{(1-q_j^i)}$}} & $D_k=log(k+1)$ & $\alpha\mbox{-}NDCG$ \\
\cline{4-5}
 & & $\textit{or simplified to}$ & $D_k=k$ & $ERR\mbox{-}IA$ \\
\cline{4-5}
 & & $Q_i^k = g_i^k(1-\alpha)^{c_j^k}$ & $D_k=(1/ \beta)^{k-1}$ & $NRBP$  \\
\hline

\end{tabular}
\end{table*}

Following the practice of machine learning, our goal is to learn a hypothesis function $h: \mathcal{X} \to \mathcal{Y}$ between an input space $\mathcal{X}$ and output space $\mathcal{Y}$. Here $\mathcal{X}$ denotes the space of possible candidate sets $\textbf{x}$, $\mathcal{Y}$ denotes the space of predicted rankings $\textbf{y}$. In order to quantify the quality of a prediction $\textbf{y}=h(\textbf{x})$, we will consider a loss function $\Delta: \mathcal{Y} \times \mathcal{Y} \to \Re $. $\Delta(\textbf{y}^{(i)}, \textbf{y})$ quantifies the penalty of prediction $\textbf{y}$ if the correct output is $\textbf{y}^{(i)}$ for given $\textbf{x}^{(i)}$.


We restrict ourselves to the supervised learning scenario. Given a set of training examples $S=\{(\textbf{x}^{(i)},\textbf{y}^{(i)}) \in \mathcal{X} \times \mathcal{Y}:i=1,...,n\}$, the learning strategy is to find a function $h$ which minimizes the empirical risk defined as:
\begin{displaymath}
R_S^{\Delta}{(h)}=\frac{1}{n} \sum_{i=1}^{n}{\Delta(\textbf{y}^{(i)},h(\textbf{x}^{(i)}))}
\end{displaymath}
In the case of learning a diverse ranking, we define the loss based on the diversity correlated evaluation measures (DCEM) as follows:
\begin{equation}
\label{eq:dcem}
\Delta_{DCEM}{(y^{(i)},y)}=1-\frac{DCEM(\textbf{y})}{DCEM(\textbf{y}^{(i)})}
\end{equation}

In this paper, we mainly consider three diversity correlated evaluation measures: ERR-IA, $\alpha$-NDCG and NRBP,
which are the current official evaluation measures of TREC diversity task \cite{TREC11}.
The corresponding diversity losses are denoted as $\Delta_{ERR\mbox{-}IA}$, $\Delta_{\alpha\mbox{-}NDCG}$ and $\Delta_{NRBP}$, respectively. Without confusion, DCEM stands for the three measures hereafter. Table~\ref{tab:measure} provides a general view of them. Their detailed explanation information can be referred to the corresponding literature \cite{comparative}.

Taking $\alpha\mbox{-}NDCG$ for example, $\alpha\mbox{-}NDCG$ is formulated as follows:
\begin{displaymath}
\alpha\mbox{-}NDCG = \frac{1}{\mathcal{N}}\sum_{i=1}^{M}p_i\sum_{k=1}^{K}\frac{g_i^k(1-\alpha)^{c_j^k}}{log_{2}(k+1)}
\end{displaymath}
where $g_i^k$ is a binary relevance value for document at postion $k$ with respect to subtopic $i$, $\alpha$ is a constant belong to $(0,1]$, $c_j^k=\sum_{j=1}^{k-1}g_i^j$, which is the number of documents ranked before position $k$ that are judged relevant to subtopic $i$, $K$ is the number of documents in a ranking list, $M$ is the number of subtopics, $p_i$ is the probability of each subtopic, and $\mathcal{N}$ is a normalization factor.

The above learning framework requires the knowledge of $\textbf{y}^{(i)}$ to use as training data. However, such $\textbf{y}^{(i)}$ are not always provided in existing public data sets. Taking the TREC diversity task \cite{TREC11} as an example, the original labeled data are provided in the form of: $(q^{(i)}, \textbf{x}^{(i)}, \mathcal{\textbf{T}}^{(i)},P(x_j^{(i)}|t):t \in \mathcal{\textbf{T}}^{(i)}, x_j^{(i)} \in \textbf{x}^{(i)})_{i=1}^{n}$, where $\textbf{x}^{(i)}$ is a candidate document set of the query $q^{(i)}$, $\mathcal{\textbf{T}}^{(i)}$ is the subtopic set of query $q^{(i)}$, $t$ is a specific subtopic in $\mathcal{\textbf{T}}^{(i)}$, and $P(x_j^{(i)}|t)$ describes the relevance of document $x_j^{(i)}$ to subtopic $t$.

Due to the non-convexity property of DCEM measures, it is NP-hard to find the optimal output $\textbf{y}^{(i)}$ with the maximal DCEM value. Therefore, we turn to a greedy selection process as described in Algorithm 1 to construct $\textbf{y}^{(i)}$, which can be viewed as an approximate optimal output. The operator $\oplus$ denotes adding a document to the already selected set.
According to the results in \cite{rishabh2012-sub,nemhauser1978analysis},
if a submodular function is monotonic (i.e., $f(S) \le f(T)$, whenever $S \subseteq T$) and normalized (i.e., $f(\phi)=0$), greedily constructing a set of size $K$ gives an $(1-1/e)$-approximation to the optimal.
Since any member of DCEM is a submodular function, we can prove that Algorithm 1 is $(1-1/e)$-approximation to the optimal (we omit the proof here). Therefore, the quality of the training data can be guaranteed in theory.

\begin{algorithm}[t]
\renewcommand{\algorithmicrequire}{	\textbf{Input:}}
\renewcommand{\algorithmicensure}{	\textbf{Output:}}
\caption {\text{Training Data Construction via Greedy Selection}}
\label{alg:greedy}
\begin{algorithmic}[1]
\REQUIRE~~\\
$(q^{(i)}, \textbf{x}^{(i)}, \mathcal{\textbf{T}}^{(i)},P(x_j^{(i)}|t):t \in \mathcal{\textbf{T}}^{(i)}, x_j^{(i)} \in \textbf{x}^{(i)})$
\ENSURE $\textbf{y}^{(i)}$
\STATE \text{Initialize solution} $\textbf{y}^{(i)} \gets \emptyset$
\FOR{$k=1,...,K$}
\STATE $bestDoc \gets \mathop{\text{argmax}}_{\{d \in \textbf{x}^{(i)} \setminus \textbf{y}^{(i)}\}}{DCEM(\textbf{y}^{(i)}\oplus d)}$
\STATE $\textbf{y}^{(i)} \gets \textbf{y}^{(i)} \cup bestDoc$
\ENDFOR
\STATE \textbf{return} $\textbf{y}^{(i)}$
\end{algorithmic}
\end{algorithm}

\section{Discriminant Function}

We focus on hypothesis function $h(\cdot;\textbf{w})$ which is parameterized by a weight vector $\textbf{w}$, and thus wish to find $\textbf{w}$ to minimize the empirical risk, $R_S^{\Delta}{(\textbf{w})} \equiv R_S^{\Delta}{(h(.;\textbf{w}))}$. Our approach is to learn a discriminant function $F: \mathcal{X} \times \mathcal{Y} \to \Re $, which can measure the quality of the predicted ranking $\textbf{y}$ for $\textbf{x}$. Given $\textbf{x}$, we can derive a prediction by finding the diverse ranking $\textbf{y}$ that maximizes $F$:
\begin{equation}
\label{eq:fun}
\begin{split}
h(\textbf{x};\textbf{w})=\mathop \text{argmax}_{y \in \mathcal{Y}}{F(\textbf{x}, \textbf{y};\textbf{w})}.
\end{split}
\end{equation}

%
A proper discriminant function should be with strong discriminative power between high quality and low quality predictions. Different retrieval settings may determine different discriminant functions.
In this section, we will try to define proper discriminant function for diverse ranking.


\subsection{Formulation of Bi-criteria Objective}

\begin{figure}
\centerline{\includegraphics[width=0.4 \textwidth]{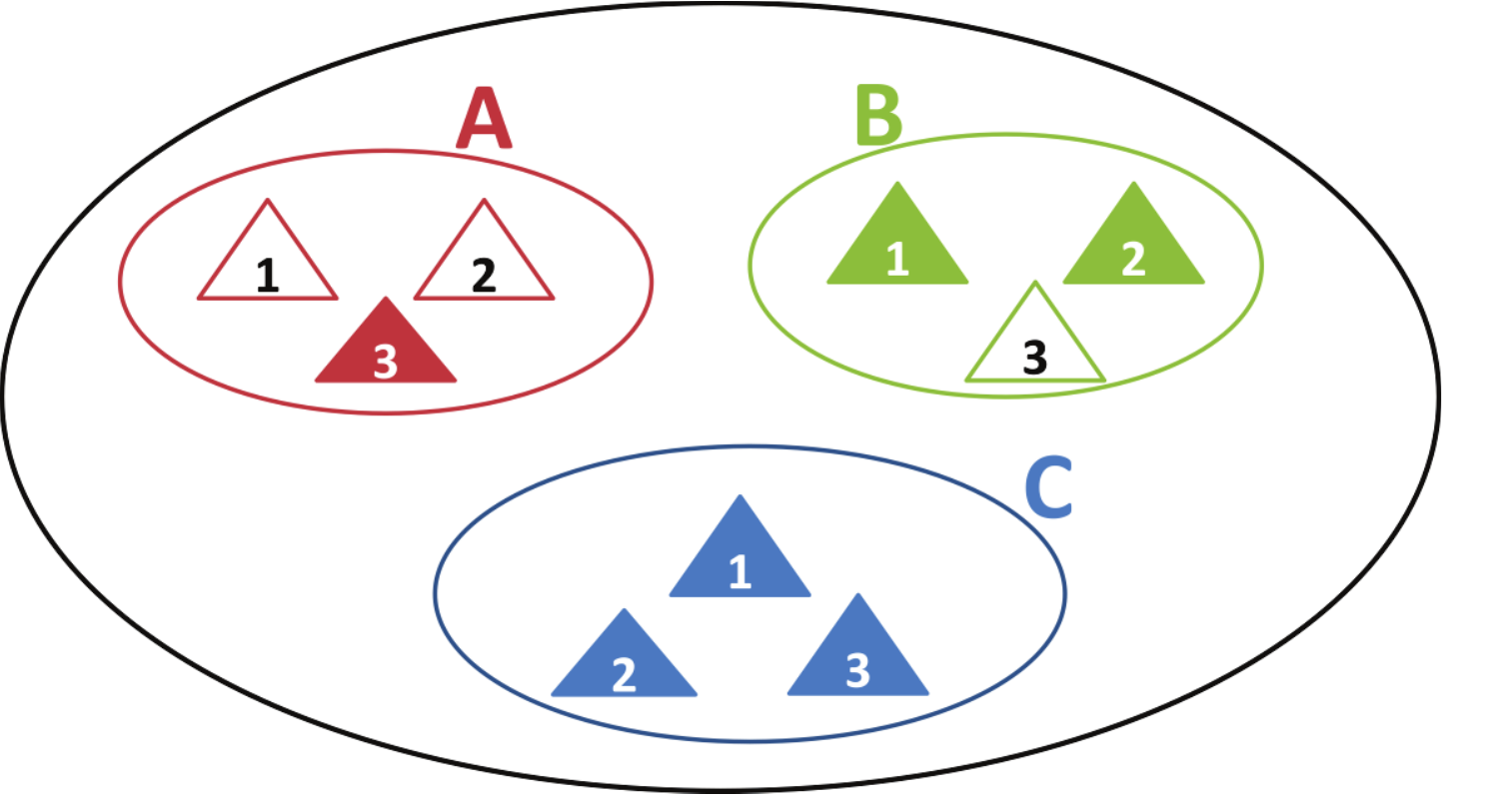}}
\caption{An Example of Ranking Prediction. All the triangles represent candidate documents $\textbf{x}^{(i)}$ of a query, and the \textit{\{A, B, C\}} sets with different colors represent different subtopics (denoted as $\mathcal{\textbf{T}}^{(i)}$). The solid triangle in each set is relevant to the user query, and the hollow triangle is irrelevant to the query.}
\label{fig:illu}
\vspace{-0.2cm}
\end{figure}

\begin{figure}
\centerline{\includegraphics[width=0.4 \textwidth]{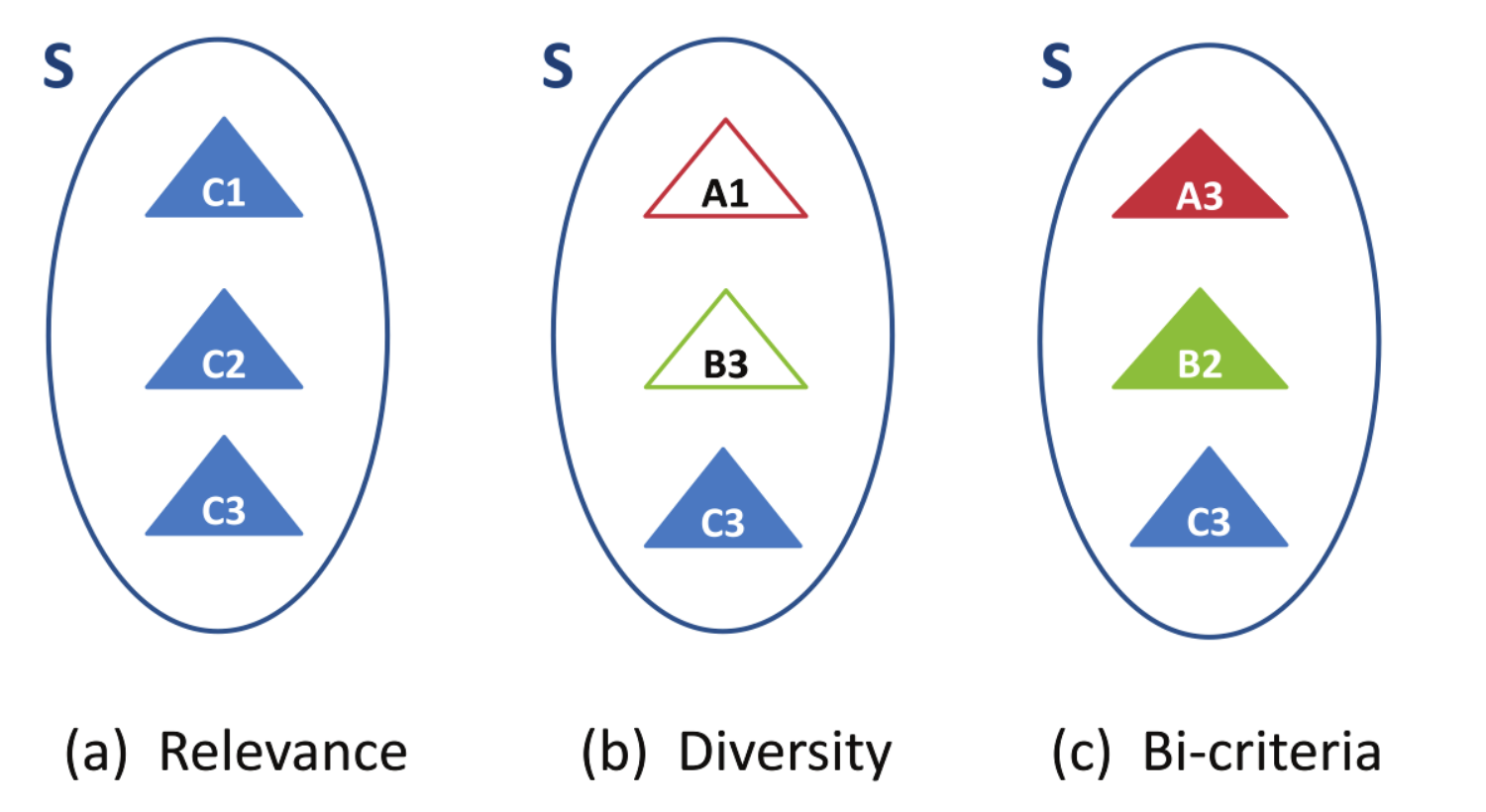}}
\caption{Solution examples for Fig.1 based on different criteria}
\label{fig:illu}
\vspace{-0.2cm}
\end{figure}

Here we first analyze our objective, the member of DCEM measures summarized in Table 1, which is the basis of our diversity loss. These measures have the same nature, and are different in some tiny components such as the way of position discounting.
We find that there are 2 key points in these measures: \textit{diversity} and the \textit{gain}. The \textit{diversity} means intent (or subtopic) coverage, which is based on explicit subtopic information of a query. Specific to a certain subtopic, the \textit{gain} describes redundancy penalizing and position discounting when accumulating the \textit{relevance} in every rank. The \textit{gain} of a specific document must be firstly based on its \textit{relevance} degree. In general, the final values of DCEM measures can be viewed as a comprehensive consideration of explicit diversity information and basic relevance score.
Therefore, we define our discriminant function as a bi-criteria objective to take both relevance and diversity into consideration.

Figure 1 is a reduced example to illustrate our prediction problem. If $\mathcal{\textbf{T}}^{(i)}$ were known, we could use Algorithm 1 to find a solution with high DCEM value. For $K=3$, the optimal solution in Figure 1 is $\textbf{y}^{(i)}=\{A_3, B_1 / B_2, C_1/C_2 / C_3\}$.

In general however, the $\mathcal{\textbf{T}}^{(i)}$ were unknown. Instead we assume that the candidate set contains a set of discriminative features that can separate subtopics from each other, and reflect the relevance degree of each document. For example, we can use topic models to model implicit topics of documents and consider distances between document pairs based on implicit topics.
If the relevance of each document is specified by a weight function $w(\cdot)$, the distance (or diversity) function between document pairs is specified as: $d(\cdot,\cdot)$, and the set selection function is denoted as: $f(\cdot)$, then a natural bi-criteria objective is to maximize the sum of relevance and dissimilarity of the selected set. It can be simply defined as follows:
\begin{equation}
\label{eq:bicri}
f(S)=\lambda_1\sum_{u \in S}{w(u)}+\lambda_2\sum_{u,v \in S}{d(u,v)}
\end{equation}
where $S$ is the solution set, and $\lambda_1>0$, $\lambda_2>0$, which are parameters for trade-off.

This type of bi-criteria objective has strong discriminative power between high and low quality predictions, which are with
high and low DCEM values, respectively.
For example, supposing all the candidate documents with binary relevance, i.e.~0 or 1, and the distance between document pairs is 1 if they belong to different subtopics, else is 0. We assign both values of $\lambda_1$ and $\lambda_2$ as 1 for simplification.
Then we can get a optimal solution with the maximal value (i.e.,~$f(S)=6$), as shown in Figure 2(c), which is the same as the optimal solution achieved based on DCEM. If we choose other solution based on \textit{sole} criterion such as relevance or diversity, the solution will be with lower value of $f(S)$, such as Figure 2(a) and Figure 2(b) (i.e.,~$f(S)=3$ and $f(S)=4$, respectively), and they are also with lower DCEM value at the same time.

In fact, the bi-criteria objective $f(S)$ shares similar insight as the work in \cite{HeTMS12}, without knowledge of this work. Despite sharing some similarities, the details of two work differ greatly, with their work mainly giving a generic theoretical analysis for a generic setting (e.g.,~properties of NP-hard, submodularity and monotonicity.), while our work presenting a structural learning framework for jointly modeling relevance and diversity based on the bi-criteria objective.

\subsection{Definition of Discriminant Function}

We assume $F$ to be linear in a combined feature representation $\Psi:\mathcal{X} \times \mathcal{Y} \to \Re^m $, which can be denoted as
\begin{equation}
\label{eq:dis}
F(\textbf{x}, \textbf{y};\textbf{w}) = \textbf{w}^{T} \Psi(\textbf{x}, \textbf{y}).
\end{equation}

As a proxy for maximizing DCEM values, we then formulate our discriminant function based on this type of bi-criteria objective (i.e., Equation~\ref{eq:bicri}) as following:
\begin{equation}
\label{eq:bidis}
\textbf{w}^{T} \Psi(\textbf{x}, \textbf{y})= \sum_{r \in \mathcal{Y}}{\textbf{w}_{r}^{T} \psi_{r}{(\textbf{x}, \textbf{y})}} + \sum_{d \in \mathcal{Y} \times \mathcal{Y}}{\textbf{w}_{d}^{T} \psi_{d}{(\textbf{x}, \textbf{y})}}
\end{equation}
where $\psi_{r}{(\textbf{x}, \textbf{y})}$ denotes the independent document feature vector describing the \textit{relevance} of a single document.
The relevance feature vector $\psi_{r}{(\textbf{x}, \textbf{y})}$ contains all the standard features traditionally adopted in the learning-to-rank literature \cite{qin:letor}, such as: the standard weighting models, field-based models, term dependence models, link analysis and URL features.
$\psi_{d}{(\textbf{x}, \textbf{y})}$ denotes the \textit{diversity} feature vector describing the dissimilarity between $d_i$ and $d_j$ pair in $\textbf{y}$, which contains a set of discriminative features that can separate subtopics from each other, in order to capture diversity effectively.
$\textbf{w}_{r}^{T}$ and $\textbf{w}_{d}^{T}$ stands for the corresponding weight vector for relevance and diversity.

Obviously, the first part of Equation~\ref{eq:bidis} is a relevance discriminant function, and the second part is a diversity discriminant function. Two kinds of discriminant functions are combined together under a bi-criteria objective in a structural learning model.

\subsection{Diversity Features}
How to define powerful features that can well capture diversity is non-trivial and critical for the success of the learning framework. Our work only provides a general direction, and presents some representative features used in our work.

\textbf{Topic Model Diversity.} Topic models, such as the probabilistic latent semantic analysis (pLSA)~\cite{Hofmann:1999}, are commonly used to model implicit topics associated with a set of documents. For diversity problem, it is necessary to associate the implicit subtopics to the documents to be ranked, and then the relations between subtopics can be converted to the relations between documents.

For a training set, we apply pLSA on the candidate sets to get the implicit subtopic distribution. Then we can define the diversity feature based on implicit topics as following:
\begin{displaymath}
\phi_{d_{topic}}{(d_i,d_j)}= \sqrt{\sum_{k=1}^{m}{(p(z_k|d_i)-p(z_k|d_j))^2}}
\end{displaymath}

\textbf{Text Diversity.} We can compute the text dissimilarity by TFIDF cosine based on vector space model (VSM), and defined as following:
\begin{displaymath}
\phi_{d_{text}}{(d_i,d_j)}=1-\frac{\textbf{d}_i \cdot \textbf{d}_j}{\Vert {\textbf{d}_i} \Vert \Vert \textbf{d}_j \Vert}
\end{displaymath}
where $\textbf{d}_i$,$\textbf{d}_j$ is the weighted document vector based on TFIDF weight.
There also exists other computing ways such as the work in \cite{Gollapudi:2009}, which is based on sketching algorithm and Jaccard similarity.

\textbf{Title Diversity.} The way of computing title diversity is the same as text diversity. We list it separately here mainly due to they belong to different document fields. We noted it as $\phi_{d_{title}}{(d_i,d_j)}$.

\textbf{Anchor Text Diversity.} The anchor text can accurately describe the content of corresponding page. Therefore, it is also an important field for a document. The way of computing anchor text diversity is the same as text and title. We noted it as $\phi_{d_{anchor}}{(d_i,d_j)}$.

\textbf{ODP-Based.} The existing ODP taxonomy\footnote{http://www.dmoz.org/} offers a succinct encoding of distances between pages (or documents). Usually, the distance between pages on similar topics in the taxonomy is likely to be small. For two categories $u$ and $v$, we define the categorical distance between them as following:
\begin{displaymath}
dis(u,v)=1-\frac{|l(u,v)|}{\max\{|u|,|v|\}}
\end{displaymath}
where $l(u,v)$ is the length of their longest common prefix. $|u|$ and $|v|$ is the length of category $u$ and $v$. For instance, given two categories: `Arts/Movies/Awards/' and `Arts/Movies/Filmmaking/Directing/Directors/', their distance is 3/5, since they share the common prefix `Arts/Movies' and the length of the longest category is 5. Then given two documents $d_i$ and $d_j$ and their category information sets $\mathcal{C}_i$ and $\mathcal{C}_j$ respectively, we define the ODP-based diversity feature as:
\begin{displaymath}
\phi_{d_{odp}}(d_i,d_j)= \frac{\sum_{u \in \mathcal{C}_i}{\sum_{v \in \mathcal{C}_j}{dis(u,v)}}}{|\mathcal{C}_i| \cdot |\mathcal{C}_j|}
\end{displaymath}
where $|\mathcal{C}_i|$ and $|\mathcal{C}_j|$ are the number of categories in corresponding category sets.

Except the semantic diversity based on the dissimilarity of document content, we also can define the diversity features from a non-semantic aspect, such as url information, or web link structure graph.

\textbf{Link-Based.} By constructing a web link graph, we can calculating the link similarity of any document pair based on direct inlink or outlink information. The link-based diversity feature is then defined as follows:
\begin{displaymath}
\phi_{d_{link}}(d_i,d_j)=
\begin{cases}
0& \text{$url_i \in {inlink(d_j) \cup outlink(d_j)}$, v.v.} \\
1&\text{other cases}
\end{cases}
\end{displaymath}

\textbf{URL-Based.} Given the url information of two documents, we can judge whether they belong to the same domain or the same site. Then we can simply define the url-based diversity feature as follows:
\begin{displaymath}
\phi_{d_{url}}(d_i,d_j)=
\begin{cases}
0&\text{one url is another's \textit{prefix}} \\
0.5&\text{belonging to the same \textit{site} or \textit{domain}} \\
1&\text{other cases}
\end{cases}
\end{displaymath}

Moreover, there also exists other useful resource information for the definition of diversity features, such as clickthrough logs.
The information of user clickthrough log is very important, and we will take it into consideration in future.

\section{Training with Structural SVM}

Structural SVM has been shown to be robust and effective, when solving complex learning problem with non-smooth ranking loss in information retrieval \cite{Tsochantaridis:2005,KDD08,Yue:2008}.
In this paper, we use structural SVM to learn the weight vector $\textbf{w}$.
\newtheorem{proposition}{Optimization Problem}
\begin{proposition}
[STRUCTURAL SVM]
\begin{equation}
\mathop {\text{min}}_{\textbf{w},\xi_i \ge 0}{\frac{1}{2} {\Vert \textbf{w}\Vert} ^2} + \frac{C}{n} \sum_{i=1}^{n}{\xi _i}
\end{equation}
$s.t. ~ \forall i,\forall \textbf{y} \in \mathcal{Y}\setminus \textbf{y}^{(i)}:$
\begin{equation}
\textbf{w}^{T} \Psi(\textbf{x}^{(i)}, \textbf{y}^{(i)}) \ge \textbf{w}^{T} \Psi(\textbf{x}^{(i)}, \textbf{y}) + \Delta_{DCEM}{(\textbf{y}^{(i)},\textbf{y})} - \xi_i
\end{equation}
\end{proposition}
The objective function (6) to be minimized is a trade-off between model complexity: ${\Vert \textbf{w} \Vert}^2$, and a hinge loss relaxation of the training loss for each training example: $\sum{\xi _i}$. In SVM training, parameter $C$ controls the trade-off and can be turned to achieve good performance for different training tasks. The $\textbf{y}^{(i)}$ is the optimal solution that can be chosen via greedy selection as the Algorithm~\ref{alg:greedy}, and $\textbf{y}^{(i)}$ minimizes $\Delta_{DCEM}{(\textbf{y}^{(i)},\textbf{y})}$, as the definition in Equation~\ref{eq:dcem}.

For each $(\textbf{x}^{(i)}, \textbf{y}^{(i)})$ in the training set, a set of constraints is added to the optimization problem as the form in Equation~(7), and the number is exponential. Despite the large number of constraints, we can employ Algorithm 2 to solve OP 1.
Algorithm 2 is a cutting plane algorithm, iteratively adding constraints until we have solved the original problem with a desired tolerance $\epsilon$ \cite{Tsochantaridis:2005}. The algorithm starts with no constraints, and iteratively finds for each training example $(\textbf{x}^{(i)}, \textbf{y}^{(i)})$, the output $\hat{\textbf{y}}$ associated with the most violated constraint. If the corresponding constraint is violated more than $\epsilon$, we add $\hat{\textbf{y}}$ into the working set $\mathcal{W}_i$ of active constraints for sample $i$, and re-solve (5) using the updated $\mathcal{W}$. It has been shown that Algorithm 2's outer loop is guaranteed to halt with a polynomial number of iterations for any desired precision $\epsilon$ \cite{Tsochantaridis:2005}.

Within the inner loop of Algorithm~2, we have to compute $\text{argmax}_{\textbf{y} \in \mathcal{Y}}{H(\textbf{y};\textbf{w})}$, where
\begin{displaymath}
H(\textbf{y};\textbf{w}) = \Delta_{DCEM}{(\textbf{y}^{(i)},\textbf{y})} + \textbf{w}^{T} \Psi(\textbf{x}^{(i)}, \textbf{y}) -  \textbf{w}^{T} \Psi(\textbf{x}^{(i)}, \textbf{y}^{(i)})
\end{displaymath}
or equivalently,
\begin{equation}
\label{eq:vio}
\mathop \text{argmax}_{\textbf{y} \in \mathcal{Y}} \Delta_{DCEM}{(\textbf{y}^{(i)},\textbf{y})} + \textbf{w}^{T} \Psi(\textbf{x}^{(i)}, \textbf{y})
\end{equation}

In fact, solving Equation~(\ref{eq:vio}) exactly is intractable, and an approximate method can be easily applied as Algorithm 1. Despite using an approximate constraint generation method, SVM training is still known to terminate in a polynomial number of iterations. Moreover, in practice, training procedure typically converges much faster than the worst case considered by the theoretical bounds \cite{Yue:2008}, and we will evaluate it empirically in the following sections.

Once the weight vector $\textbf{w}$ is obtained, the prediction procedure can be made via Equation~\ref{eq:dis} by employing a greedy selection approach as Algorithm~1, with using $\textbf{w}^{T} \Psi(\textbf{x}^{(i)}, \textbf{y})$ to replace the corresponding DCEM measure, and iteratively selecting the document with the highest marginal gain.

\begin{algorithm}[tb]
\renewcommand{\algorithmicrequire}{	\textbf{Input:}}
\renewcommand{\algorithmicensure}{	\textbf{Output:}}
\caption {Cutting Plane Algorithm for Solving OP 1 with tolerance $\epsilon$}
\label{alg:pf}
\begin{algorithmic}[1]
\REQUIRE~{$(\textbf{x}^{(1)},\textbf{y}^{(1)})$,...,$(\textbf{x}^{(n)},\textbf{y}^{(n)})$, $C$, $\epsilon$}
\STATE {$\mathcal{W}_i \gets \emptyset $ \text{for all} $i=1,..,n$}
\REPEAT
\FOR{$i=1,...,n$}
\STATE $H(\textbf{y};\textbf{w}) \equiv \Delta_{DCEM}{(\textbf{y}^{(i)},\textbf{y})} + \textbf{w}^{T} \Psi(\textbf{x}^{(i)}, \textbf{y}) -  \textbf{w}^{T} \Psi(\textbf{x}^{(i)}, \textbf{y}^{(i)})$
\STATE \text{compute} $\hat{\textbf{y}}=\text{argmax}_{y \in \mathcal{Y}}H(\textbf{y};\textbf{w})$
\STATE \text{compute} $\xi _i=\text{max}\{0,\text{max}_{y \in \mathcal{W}_i}H(\textbf{y};\textbf{w})\}$
\IF{$H(\hat{\textbf{y}}; \textbf{w}) > \xi _i + \epsilon$}
\STATE $\mathcal{W}_i \gets \mathcal{W}_i \cup \{\hat{\textbf{y}}\}$
\STATE $\textbf{w} \gets \text{optimize (5) over}~\mathcal{W}=\bigcup_i \mathcal{W}_i$
\ENDIF
\ENDFOR
\UNTIL {\text{no}~$\mathcal{W}_i$~\text{has changed during iteration}}
\end{algorithmic}
\end{algorithm}

\section{Experiments}

In this section, we evaluate the effectiveness of our approach empirically. In particular, we compare against a series of popular diversification approaches using official TREC diversity measures and traditional diversity measures. Furthermore, we analyze the performance robustness of different diversification approaches.
In addition, we study the effect of our approximate constraint generation in training procedure, and analyze the importance of our proposed diversity features. Finally we study the efficiency of our approach based on the analysis of running time.

\subsection{Experimental Setup}
Here we give some introductions on the experimental setup, including data collections, evaluation metrics, baseline models and experiment design.

\textbf{Data Collections.} Our experiments are conducted in the context of the diversity task of the TREC2009 Web Track (WT2009), TREC2010 Web Track (WT2010), and TREC2011 Web Track (WT2011), which contain 50, 48 and 50 test queries (or topics), respectively. Each topic includes several subtopics identified by TREC assessors, with binary relevance judgements provided at the subtopic level\footnote{In fact, for WT2011 task, assessors made graded judgements. While in the official TREC evaluation program, it mapped these graded judgements to binary judgements by treating $values>0$ as relevant and $values \le 0$ as not relevant.}. Our evaluation is done on the ClueWeb09 Category B data collection\footnote{http://boston.lti.cs.cmu.edu/Data/clueweb09/}, which comprises a total of 50 million English Web documents.

\textbf{Evaluation Metrics.} The current official evaluation metrics of the diversity task include ERR-IA, $\alpha$-NDCG and NRBP.
They implement a cascade user model which penalizes redundancy by assuming an increasing probability that users will stop inspecting the results as they find their desired information. Additionally, we also use traditional diversity measures for evaluation, i.e., Precision-IA and Subtopic recall. They measure the precision across all subtopics of the query and the ratio of the subtopics covered in the results, respectively. All the measures are computed at rank cutoff: 20. Moreover, the associated parameters $\alpha$ and $\beta$ are all set to be 0.5, which is consistent with the default settings in official TREC evaluation program.

\textbf{Baseline Models.}
To evaluate the performance of our approach, we compare our approach with the state-of-the-art approaches, which are introduced as follows.
\begin{itemize}
\item \textbf{QL}.  The standard Query-Likelihood language model is used for conducting the initial retrieval, which provides the top 1000 retrieved documents as a candidate set for all the diversification approaches. It is also chosen as a basic baseline method in our experiment.

\item \textbf{MMR}. MMR is a classical implicit diversity method in the diversity research. It employs a linear combination of relevance and diversity as the metric called ``marginal relevance'' \cite{MMR}. MMR will iteratively select document with the largest ``marginal relevance''.

\item \textbf{xQuAD}. The explicit diversification approaches are popular in current research field, in which xQuAD is the most representative and used as a baseline model in our experiments \cite{Santos:2010}.

\item \textbf{PM-2}. PM-2 is also a explicit method that proposes to optimize proportionality for search result diversification \cite{proportionality}. It has been proved to achieve promising performance in their work, and is also chosen as a baseline method in our experiment.

\item \textbf{ListMLE}. ListMLE is a plain learning-to-rank approach without diversification considerations, yet it is the state-of-the-art listwise relevance approach in LTR field \cite{Xia:20086}. We use it as a basic supervised baseline.

\item \textbf{SVMDIV}. SVMDIV is a representative supervised approach for search result diversification \cite{Yue:2008}. It proposes to optimize subtopic coverage by maximizing word coverage. It formulates the learning problem and derives a training method based on structural SVMs. However, SVMDIV only models diversity and discards the requirement of relevance. For fair performance comparison, we will firstly apply ListMLE to do the initial ranking to capture relevance, and then use SVMDIV to re-rank top-$K$ retrieved documents to capture diversity.
\end{itemize}
In fact, the above three diversity baselines: MMR, xQuAD and PM-2, all require a prior relevance function to implement their diversification steps. In our experiment, we choose QL as the relevance function for them, and obtain three \textit{unsupervised-relevance} versions of diversification baselines: MMR$_{QL}$, xQuAD$_{QL}$ and PM-2$_{QL}$, respectively. Meanwhile, we also apply ListMLE as the relevance function to implement them, and obtain three \textit{supervised-relevance} versions: MMR$_{list}$, xQuAD$_{list}$ and PM-2$_{list}$, respectively.

With different evaluation metrics used as the objectives, our SVM$_{DCEM}$ approach has 3 variants as described in section 3, denoted as: SVM$_{ERR\mbox{-}IA}$, SVM$_{\alpha\mbox{-}NDCG}$ and SVM$_{NRBP}$, respectively.

\begin{table}[t]
\centering
\caption{Relevance Features for learning on ClueWeb09-B collection \cite{qin:letor,Metzler:2005:MRF}.}
\label{tab:relevance}
{
\addtolength{\tabcolsep}{8pt}
\begin{tabular}{c|c|c} \hline
Category & Feature Description & Total \\ \hline
$Q\mbox{-}D$ & TF-IDF & 5 \\ \hline
$Q\mbox{-}D$ & BM25 & 5 \\ \hline
$Q\mbox{-}D$ & QL.DIR & 5 \\ \hline
$Q\mbox{-}D$ & MRF & 10 \\ \hline
$D$ & PageRank & 1 \\ \hline
$D$ & Inlink number & 1 \\ \hline
$D$ & Outlink number & 1 \\ \hline
\end{tabular}}
\vspace{-0.2cm}
\end{table}

\textbf{Experiment Design.} In our experiments, we use Indri toolkit (version 5.2)\footnote{http://lemurproject.org/indri} as the retrieval platform. For the test query set on each dataset, we use a 5-fold cross validation with a ratio of 3:1:1, for training, validation and testing. The final test performance is reported as the average over all the folds.

For data preprocessing, we apply Porter stemmer and stopwords removing for indexing and query processing.
We then extract features for each dataset as follows.
For relevance, we use several standard features in learning-to-rank research \cite{qin:letor}, such as typical weighting models (e.g., TF-IDF, BM25, LM), and term dependency model (e.g., MRF), as summarized in Table~\ref{tab:relevance}, where $Q\mbox{-}D$ means that the feature is dependent on both the query and the document, $D$ means that the feature only depends on the document. For all the $Q\mbox{-}D$ features, they are applied in five fields: body, anchor, title, URL and whole document. Additionally, the MRF has two types of values: ordered phrase and unordered phrase \cite{Metzler:2005:MRF}, so the total features number is 10.
For diversity, we use both semantic and non-semantic diversity features described before (e.g., $\phi_{d_{topic}}$, $\phi_{d_{text}}$, $\phi_{d_{title}}$, $\phi_{d_{anchor}}$, $\phi_{d_{odp}}$, $\phi_{d_{link}}$, $\phi_{d_{url}}$).
For the sake of efficiency, we only consider the top 100 values of each type of diversity feature for each document, and the other values are set to be zero. Finally, all feature values are normalized to the range of [0,1].

For three baseline models: MMR, xQuAD and PM-2, they all have a single parameter $\lambda$ to tune. We perform a 5-fold cross validation to train $\lambda$ through optimizing $ERR\mbox{-}IA$. Additionally, for xQuAD and PM-2, the official subtopics are used as a representation of taxonomy classes to simulate their best-case scenarios, and uniform probability for all subtopics is assumed, as described in their work \cite{Santos:2010,proportionality}.

For ListMLE, SVMDIV and our approach, we utilize the same training data generated by Algorithm~1 (where $ERR\mbox{-}IA$ is chosen as the corresponding DCEM measure for optimizing), and conduct 5-fold cross validation.  ListMLE adopts the relevance features summarized in Table~\ref{tab:measure}. SVMDIV adopts the representative word level features with different importance criteria, as listed in their paper and released code \cite{Yue:2008}. As described in above subsection, SVMDIV will re-rank top-$K$ retrieved documents returned by ListMLE. We test $K \in \{30, 50, 100\}$, and find it performs best at $K=30$. Therefore, the following results of SVMDIV are achieved with $K=30$.

For SVMDIV and our SVM$_{DCEM}$, the $C$ parameter of the SVM is varied from $10^{-4}$ to $10^3$. The best $C$ value is chosen based on the performance of validation set.

\subsection{Performance Comparison}

\begin{table*}
\centering
\caption{Performance comparison of all methods in official TREC diversity measures for WT2009. The numbers in the parentheses are the relative improvements compared with the baseline method QL. Boldface indicates the highest scores among all runs.}
\label{tab:2009}
{
\begin{tabular}{c|ccc} \hline
&  ERR-IA & $\alpha$-NDCG & NRBP \\ \hline
QL	&	0.1637 &    0.2691 &	0.1382 \\ \hline
MMR$_{QL}$	&	0.1625~~(-0.73\%) &	0.2658~~(-1.23\%) &	0.1361~~(-1.52\%)  \\
xQuAD$_{QL}$ &	0.1922~~(+17.41\%) &	0.3093~~(+14.94\%) &	0.1674~~(+21.13\%) \\
PM-2$_{QL}$ &  0.1835~~(+12.10\%) &	   0.2896~~(+7.62\%) &	0.1607~~(+16.28\%) \\ \hline
ListMLE & 0.1913~~(+16.86\%)&	0.3074~~(+14.23\%)&	0.1681~~(+21.64\%) \\
MMR$_{list}$ & 0.2022~~(+23.52\%) &	0.3083~~(+14.57\%) &	0.1615~~(+16.86\%) \\
xQuAD$_{list}$ & 0.2316~~(+41.48\%) & 0.3437~~(+27.72\%)&	0.1956~~(+41.53\%) \\
PM-2$_{list}$ & 0.2294~~(+40.13\%)&	0.3369~~(+25.20\%)&	0.1788~~(+29.38\%) \\
SVMDIV & 0.2408~~(+47.10\%)&	0.3526~~(+31.03\%)&	0.2073~~(+50.00\%) \\ \hline
SVM$_{ERR\mbox{-}IA}$ & \textbf{0.2613}~~(+59.62\%) &	0.3726~~(+38.46\%) &	0.2195~~(+58.83\%) \\
SVM$_{\alpha\mbox{-}NDCG}$ & 0.2597~~(+58.64\%) &	\textbf{0.3765}~~(+39.91\%) &	0.2192~~(+58.61\%) \\
SVM$_{NRBP}$ & 0.2589~~(+58.16\%) &	0.3712~~(+37.94\%) &	\textbf{0.2223}(+60.85\%) \\
 \hline
 \end{tabular}}
 \vspace{-0.2cm}
\end{table*}

\begin{table*}
\centering
\caption{Performance comparison of all methods in official TREC diversity measures for WT2010. The numbers in the parentheses are the relative improvements compared with the baseline method QL. Boldface indicates the highest scores among all runs.}
\label{tab:2010}
{
\begin{tabular}{c|ccc} \hline
&  ERR-IA & $\alpha$-NDCG & NRBP \\ \hline
QL	&	0.198&	0.3024&	0.1549 \\ \hline
MMR$_{QL}$	&	0.2062~~(+4.14\%)&	0.3150~~(+4.17\%)&	0.1647~~(+6.33\%)  \\
xQuAD$_{QL}$ &	0.2583~~(+30.45\%)&	0.3882~~(+28.37\%)&	0.2160~~(+39.44\%) \\
PM-2$_{QL}$ &  0.2579~~(+30.25\%)&	0.3907~~(+29.20\%)&	0.2166~~(+39.83\%) \\ \hline
ListMLE & 0.2436~~(+23.03\%) &	0.3755~~(+24.17\%)&	0.1949~~(+25.82\%) \\
MMR$_{list}$ & 0.2735~~(+38.13\%)&	0.4036~~(+33.47\%)&	0.2252~~(+45.38\%) \\
xQuAD$_{list}$ & 0.3278~~(+65.56\%)&	0.4445~~(+46.99\%)&	0.2872~~(+85.41\%) \\
PM-2$_{list}$ & 0.3296~~(+66.46\%)&	0.4478~~(+48.08\%)&	0.2901~~(+87.28\%) \\
SVMDIV & 0.3331~~(+68.23\%)	&0.4593~~(+51.88)&	0.2934~~(+89.41\%) \\ \hline
SVM$_{ERR\mbox{-}IA}$ & \textbf{0.3546}~~(+79.09\%)&	0.4723~~(+56.18\%)&	0.3097~~(+99.94\%) \\
SVM$_{\alpha\mbox{-}NDCG}$ & 0.3521~~(+77.83\%)	&\textbf{0.4764}~~(+57.54\%)&	0.3086~~(+99.23\%) \\
SVM$_{NRBP}$ & 0.3514~~(+77.47\%)&	0.4718~~(+56.02\%)&	\textbf{0.3116}~~(+101.16\%) \\
 \hline
 \end{tabular}}
  \vspace{-0.2cm}
\end{table*}

\begin{table*}
\centering
\caption{Performance comparison of all methods in official TREC diversity measures for WT2011. The numbers in the parentheses are the relative improvements compared with the baseline method QL. Boldface indicates the highest scores among all runs.}
\label{tab:2011}
{
\begin{tabular}{c|ccc} \hline
&  ERR-IA & $\alpha$-NDCG & NRBP \\ \hline
QL	&	0.3520&	0.4531&	0.3123 \\ \hline
MMR$_{QL}$	&	0.3534~~(+0.40\%)&	0.4612~~(+1.79\%)&	0.3205~~(+2.63\%)  \\
xQuAD$_{QL}$ &	0.4231~~(+20.20\%)&	0.5268~~(+16.27\%)&	0.3991~~(+27.79\%) \\
PM-2$_{QL}$ &  0.4319~~(+22.70\%)&	0.5334~~(+17.72\%)&	0.4062~~(+30.07\%) \\ \hline
ListMLE & 0.4172~~(+18.52\%)&	0.5169~~(+14.08\%)&	0.3887~~(+24.46\%) \\
MMR$_{list}$ & 0.4284~~(+21.70\%)&	0.5302~~(+17.02\%)&	0.3913~~(+25.30\%) \\
xQuAD$_{list}$ & 0.4753~~(+35.03\%)&	0.5645~~(+24.59\%)&	0.4274~~(+36.86\%) \\
PM-2$_{list}$ & 0.4873~~(+38.44\%)&	0.5786~~(+27.70\%)&	0.4318~~(+38.26\%) \\
SVMDIV & 0.4898~~(+39.15\%)&	0.5910~~(+30.43\%)	&0.4475~~(+43.29\%) \\ \hline
SVM$_{ERR\mbox{-}IA}$ & \textbf{0.5132}~~(+45.80\%)&	0.6137~~(+35.44\%)&	0.4683~~(+49.95\%) \\
SVM$_{\alpha\mbox{-}NDCG}$ & 0.5116~~(+45.34\%)&	\textbf{0.6173}~~(+36.24\%)&	0.4679~~(+49.82\%) \\
SVM$_{NRBP}$ & 0.5112~~(+45.23\%)&	0.6129~~(+35.27\%)&	\textbf{0.4691}~~(+50.21\%) \\
 \hline
 \end{tabular}}
\vspace{-0.2cm}
\end{table*}

We now compare our approaches to the baseline models on search result diversification.
The results of performance comparison are shown in Table~\ref{tab:2009},\ref{tab:2010},\ref{tab:2011}, and we have the following observations.

(1) Regarding the comparison among representative implicit and explicit approaches, explicit methods (i.e.~xQuAD and PM-2) show better performance than the implicit method (i.e.~MMR) in terms of all the evaluation measures. MMR is the least effective due to its simple predefined ``marginal relevance'', which tries to capture novelty only based on inter-document similarity. The two explicit methods achieve comparable performance: PM-2$_{list}$ wins on WT2010 and WT2011, while xQuAD$_{list}$ wins on WT2009, but their overall performance differences are small.

Besides, for all these methods, the supervised-relevance versions (i.e.~MMR$_{list}$, xQuAD$_{list}$ and PM-2$_{list}$) are all superior than their corresponding unsupervised-relevance versions (i.e.~MMR$_{QL}$, xQuAD$_{QL}$ and PM-2$_{QL}$). The results indicate the importance of the prior relevance function, which is required for implementing their diversification steps. By learning a better relevance function, one can achieve better performance in diversification. In fact, even pure supervised relevance method ListMLE, can achieve comparable performance with the explicit methods under unsupervised-relevance versions (i.e.~xQuAD$_{QL}$ and PM-2$_{QL}$), which further proves the importance of a proper relevance function even in a diversification scenario.

(2) Learning-based methods (i.e.~SVMDIV and SVM$_{DCEM}$) further outperform the the state-of-the-art explicit methods in terms of all the evaluation measures. For example, with the evaluation of $ERR\mbox{-}IA$, the relative improvement of SVM$_{\alpha\mbox{-}NDCG}$ over the xQuAD$_{list}$ is up to 17.16\%, 12.27\%, 10.31\%, on WT2009, WT2010, WT2011, respectively, and the relative improvement of SVM$_{\alpha\mbox{-}NDCG}$ over the PM-2$_{list}$ is up to 18.51\%, 11.37\%, 6.9\% on WT2009, WT2010, WT2011, respectively.
Although xQuAD$_{list}$ and PM-2$_{list}$ all utilize the official subtopics as explicit query aspects to simulate their best-case scenarios, their performances are still much lower than learning-based approaches, which indicates that there might be certain gap between their predefined utility functions and the final evaluation measures.

(3) Comparing with the learning-based methods, our SVM$_{DCEM}$ approaches all outperform the SVMDIV method. The relative improvement of SVM$_{\alpha\mbox{-}NDCG}$ over the SVMDIV is up to 11.54\%, 9.6\%, 6.19\%, in terms of $ERR\mbox{-}IA$ on WT2009, WT2010, WT2011, respectively. We further validate that all these improvements are statistically significant ($p\mbox{-}value<0.01$).
As we know, SVMDIV simply uses weighted word coverage as a proxy for explicitly covering subtopics, while our SVM$_{DCEM}$
jointly modeling relevance and diversity based on a proper bi-criteria objective.
Therefore, our SVM$_{DCEM}$ approach shows better formulation of diverse ranking, and leads to better performance in search result diversification.

(4) Not surprisingly, the method optimizing an evaluation metric leads to the best performance with respect to the corresponding evaluation metric. For example, SVM$_{ERR\mbox{-}IA}$ performs best with $ERR\mbox{-}IA$ as the evaluation measure, SVM$_{\alpha\mbox{-}NDCG}$ performs best with $\alpha\mbox{-}NDCG$ as the evaluation measure, and SVM$_{NRBP}$ performs best with $NRBP$ as the evaluation measure, in all the three datasets, and the results are accordance with our intuition.

In addition, we also evaluate these diversity methods in traditional diversity measures: Precision-IA and Subtopic recall, and experimental results are shown in Fig.~3 and 4.
We can see that our approaches outperform all the baseline models in all the datasets, which is consistent with the evaluation results in Table~\ref{tab:2009},\ref{tab:2010},\ref{tab:2011}. When comparing the 3 variants of SVM$_{DCEM}$ approach, SVM$_{ERR\mbox{-}IA}$ and SVM$_{\alpha\mbox{-}NDCG}$ perform a little better than SVM$_{NRBP}$, yet their overall performance differences are small.

\begin{figure*}[t]
 \centering
  \subfigure{
    \label{fig:subfig:a} 
    \includegraphics[width=0.33 \textwidth]{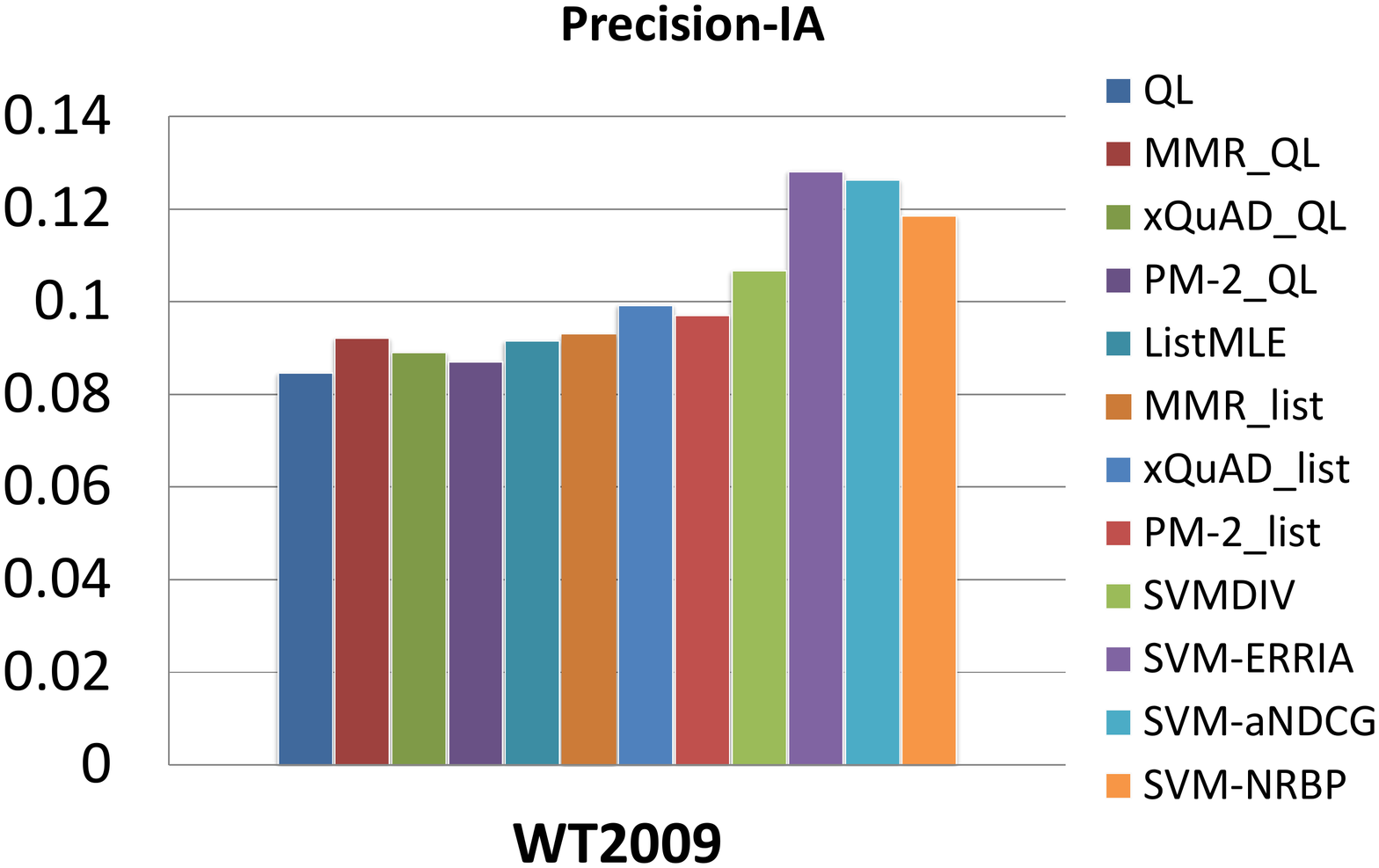}}
  \hspace{-0.3cm}
  \subfigure{
    \label{fig:subfig:a} 
    \includegraphics[width=0.33 \textwidth]{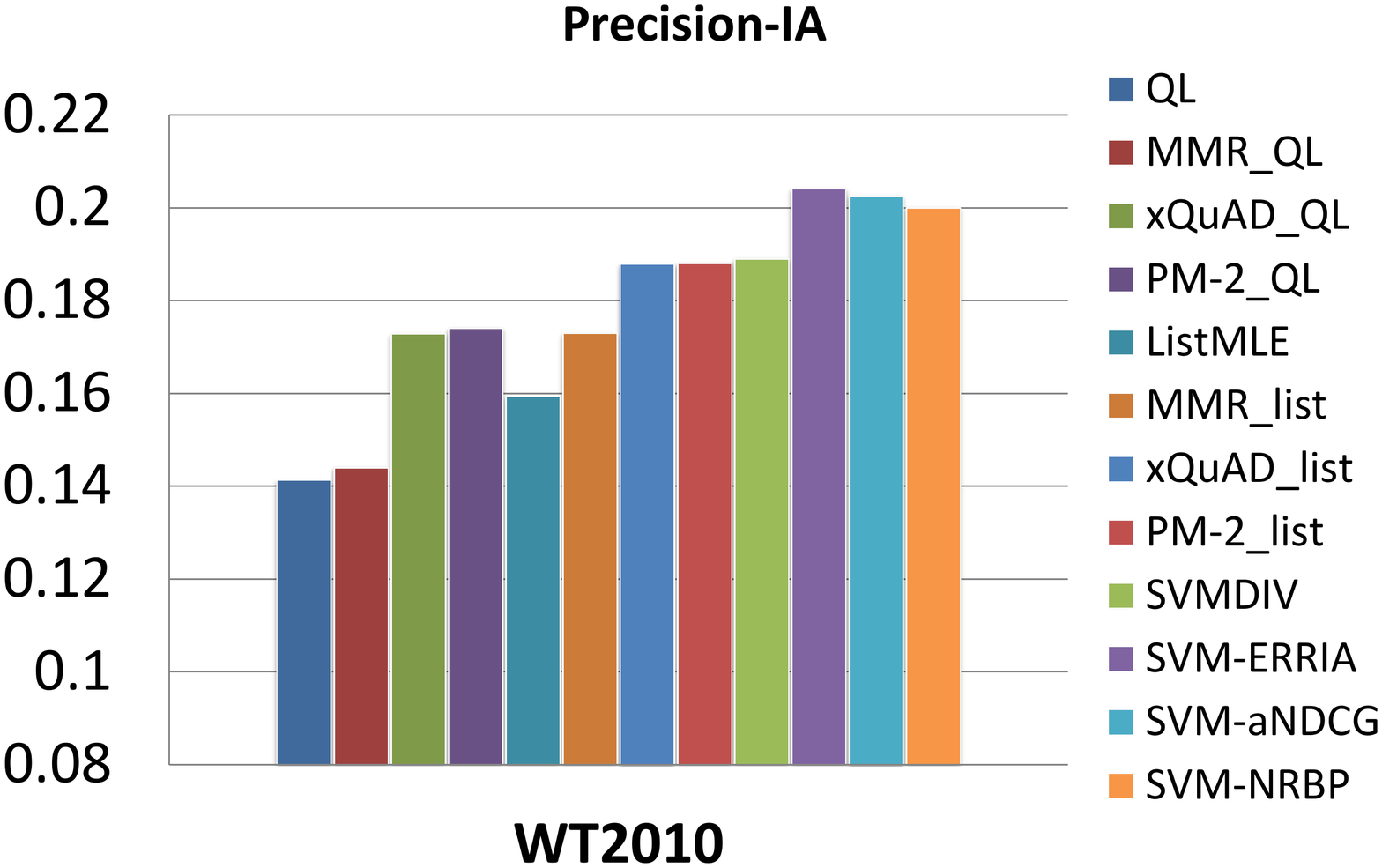}}
  \hspace{-0.3cm}
  \subfigure{
    \label{fig:subfig:b} 
    \includegraphics[width=0.33 \textwidth]{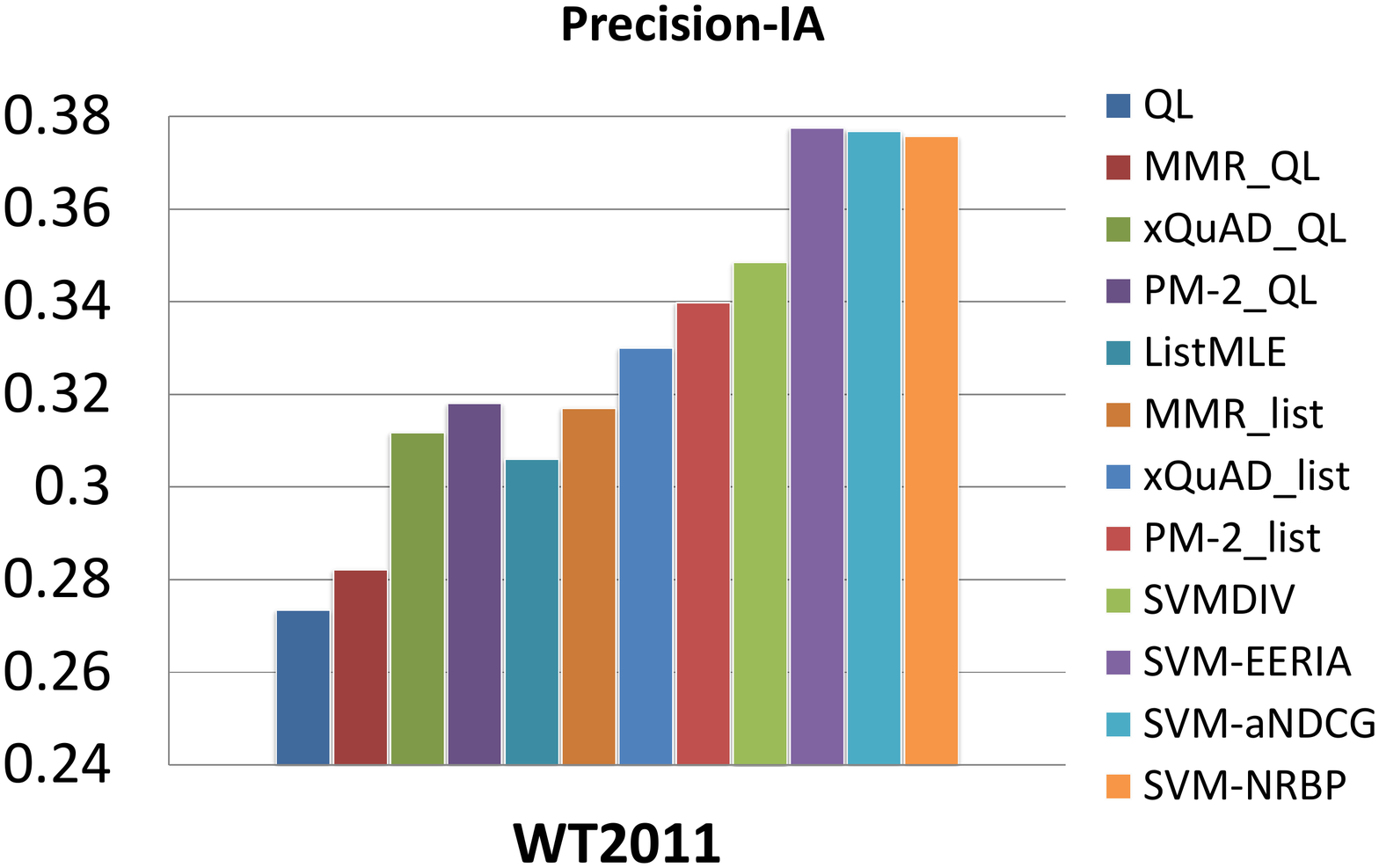}}
  \caption{Performance comparison of all methods in Precision-IA for WT2009, WT2010 and WT2011.}
  \vspace{-0.2cm}
\end{figure*}

\begin{figure*}[t]
 \centering
  \subfigure{
    \label{fig:subfig:a} 
    \includegraphics[width=0.33 \textwidth]{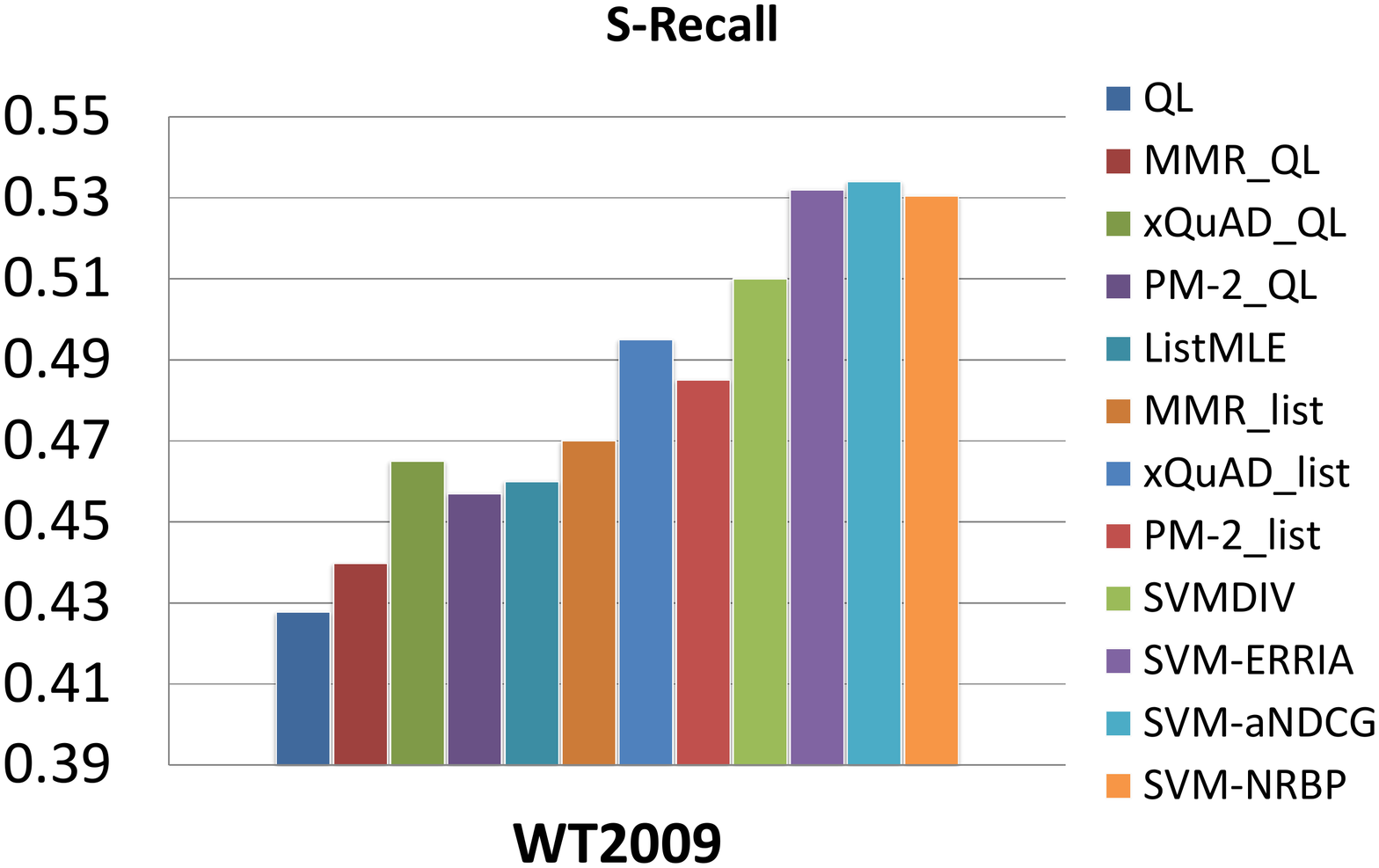}}
  \hspace{-0.3cm}
  \subfigure{
    \label{fig:subfig:a} 
    \includegraphics[width=0.33 \textwidth]{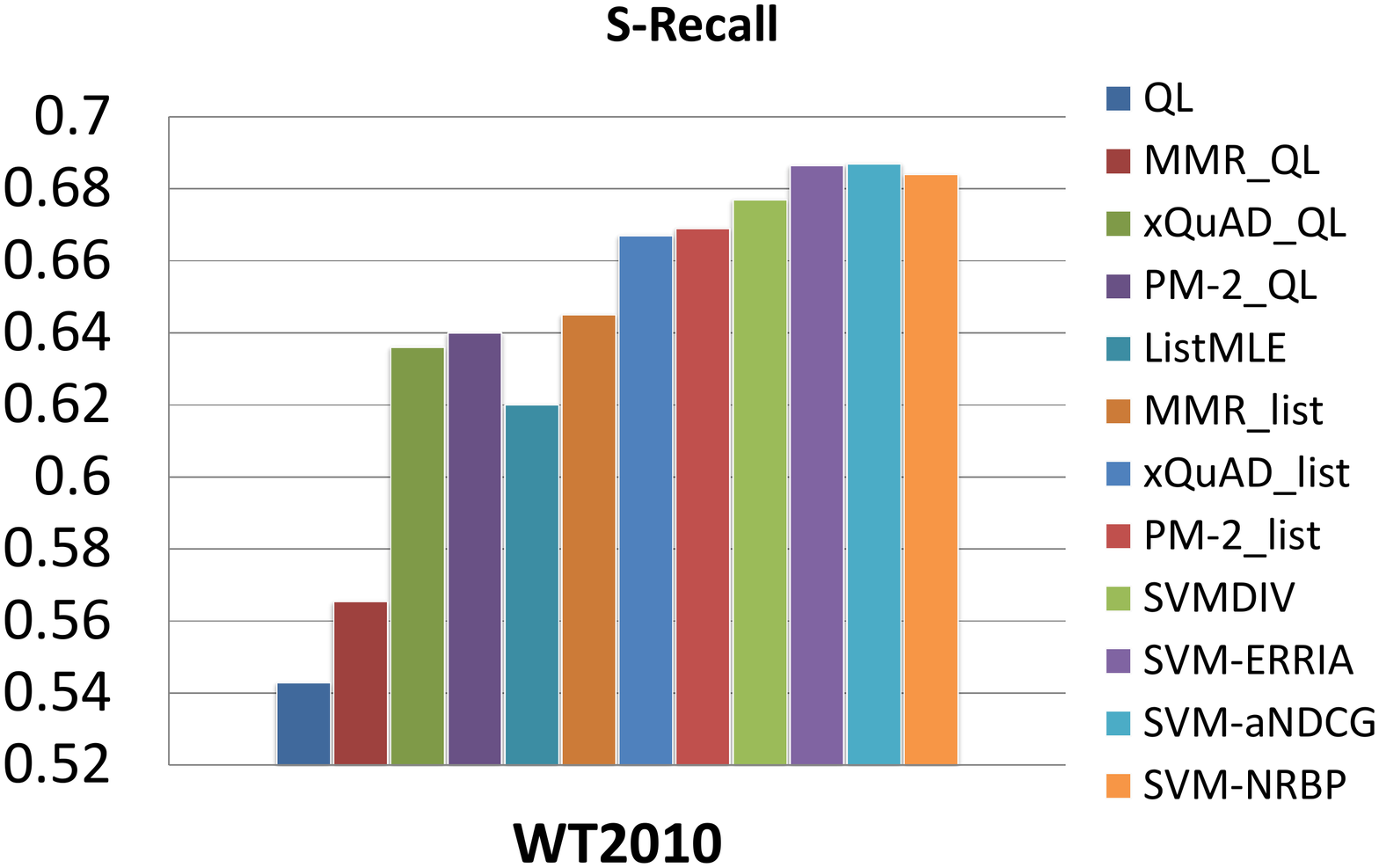}}
  \hspace{-0.3cm}
  \subfigure{
    \label{fig:subfig:b} 
    \includegraphics[width=0.33 \textwidth]{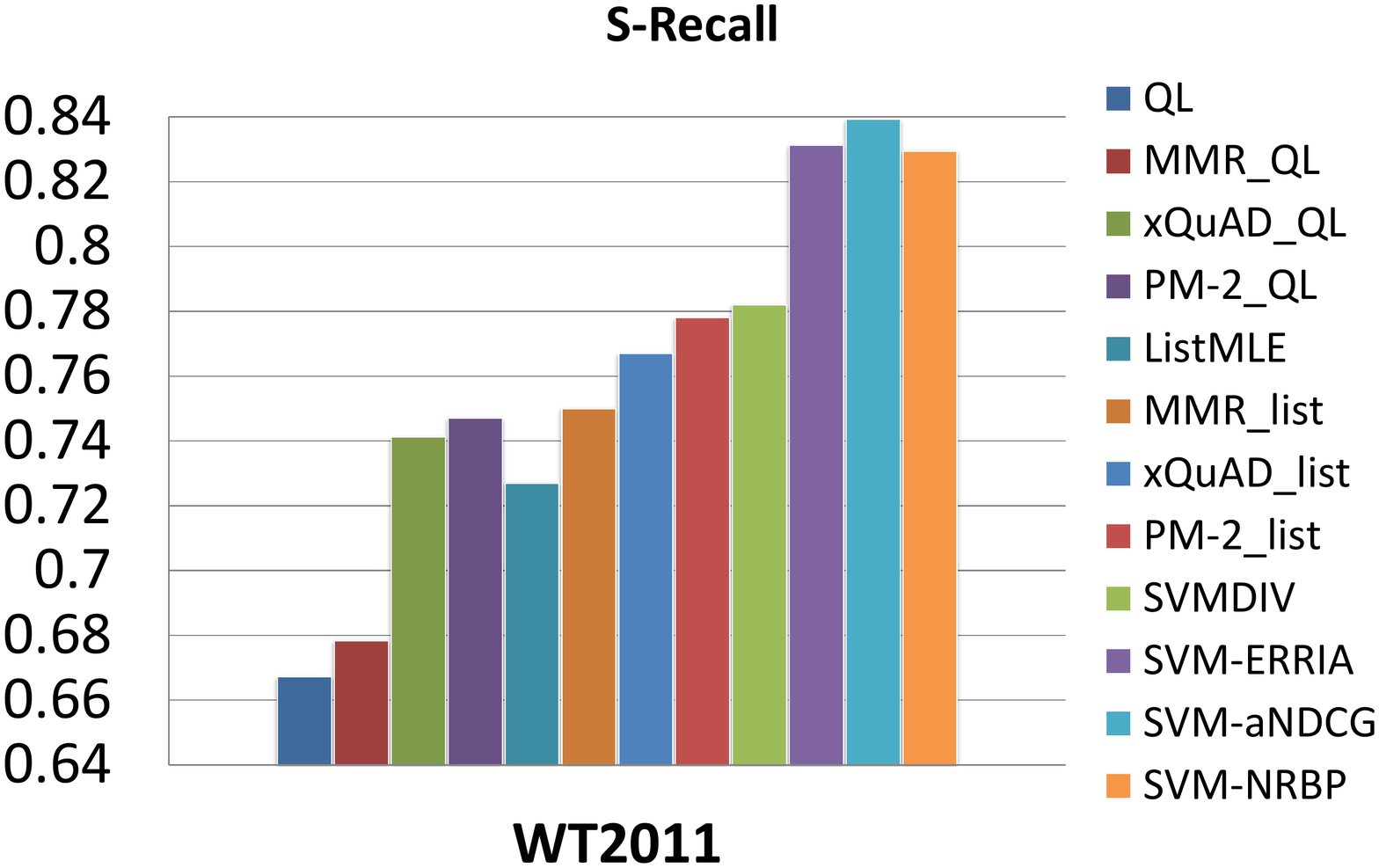}}
  \caption{Performance comparison of all methods in Subtopic Recall for WT2009, WT2010 and WT2011.}
  \vspace{-0.2cm}
\end{figure*}

\subsection{Robustness Analysis}

\begin{table}
\caption{The robustness of the performance of all diversification methods in Win/Loss ratio.}
\label{tab:robust}
{
\begin{tabular}{c|cccc} \hline
& WT2009 & WT2010 & WT2011 & \textit{Total} \\ \hline
MMR$_{QL}$ & 18/20 & 21/15 & 19/17 & 58/52 \\
xQuAD$_{QL}$ & 25/16 & 29/16 & 28/11 & 83/38 \\
PM-2$_{QL}$ & 18/19 & 30/16 & 30/12 & 78/47 \\ \hline
ListMLE & 20/18 & 27/16 & 26/11 & 73/45 \\
MMR$_{list}$ & 22/15 & 29/13 & 29/10 & 80/38 \\
xQuAD$_{list}$ & 28/11 & 31/12 & 31/12 & 90/35 \\
PM-2$_{list}$ & 26/15 & 32/12 & 32/11 & 90/38 \\
SVMDIV & 30/12 & 32/11 & 32/11 & 94/34 \\ \hline
SVM$_{ERR\mbox{-}IA}$ & 33/10 & 34/11 & 33/10 & 100/31\\
SVM$_{\alpha\mbox{-}NDCG}$ & 33/10 & 32/9 & 34/11 & 99/30 \\
SVM$_{NRBP}$ & 32/11 & 33/10 & 34/11 & 99/32 \\ \hline
\end{tabular}}
\end{table}

In this section we analyze the robustness of these diversification methods, i.e.~whether the performance improvement is consistent as compared with the basic relevance baseline model \cite{Wang:2012:RRM}. Specifically, we define the robustness as the Win/Loss ratio \cite{Yue:2008,proportionality} - the ratio of queries whose performance improves or hurts as compared with the original results from QL in terms of of $ERR\mbox{-}IA$.

From results in Table~\ref{tab:robust}, we first notice that for the implicit and explicit methods, their supervised-relevance versions (i.e.~MMR$_{list}$, xQuAD$_{list}$ and PM-2$_{list}$) show better robustness than their corresponding unsupervised-relevance versions, which is also consistent with the evaluation results in Table~\ref{tab:2009},\ref{tab:2010},\ref{tab:2011}. xQuAD performs better than PM-2 no matter supervised-relevance version or unsupervised-relevance version. Among all the diversification baselines, SVMDIV shows the best performance robustness with the total Win/Loss ratio around 2.8.
Finally, our SVM$_{DCEM}$ methods achieve the best robustness as compared with all the baseline methods, with the total Win/Loss ratio around 3.2. Among the three variants of SVM$_{DCEM}$, SVM$_{\alpha\mbox{-}NDCG}$ performs a little better than the two others, with the Win/Loss ratio as 3.3.

Based on the robustness results, we can see that the performance of our SVM$_{DCEM}$ methods are more stable than all the baseline methods. It demonstrates that the overall performance gains of our approach not only come from some small subset of queries. In other words, the result diversification for different queries could be well addressed in our approach.

\subsection{Approximate Constraint Generation}

In our work, we use an approximate way of constraint generation for model training, which may compromise our models' ability to fit the data. Similar to the study in \cite{Yue:2008}, we address this concern by examining the training loss as $C$ is varied.
A high value of $C$ indicates the training model favors low training loss over low model complexity.

\begin{figure}
\centerline{\includegraphics[width=0.45 \textwidth]{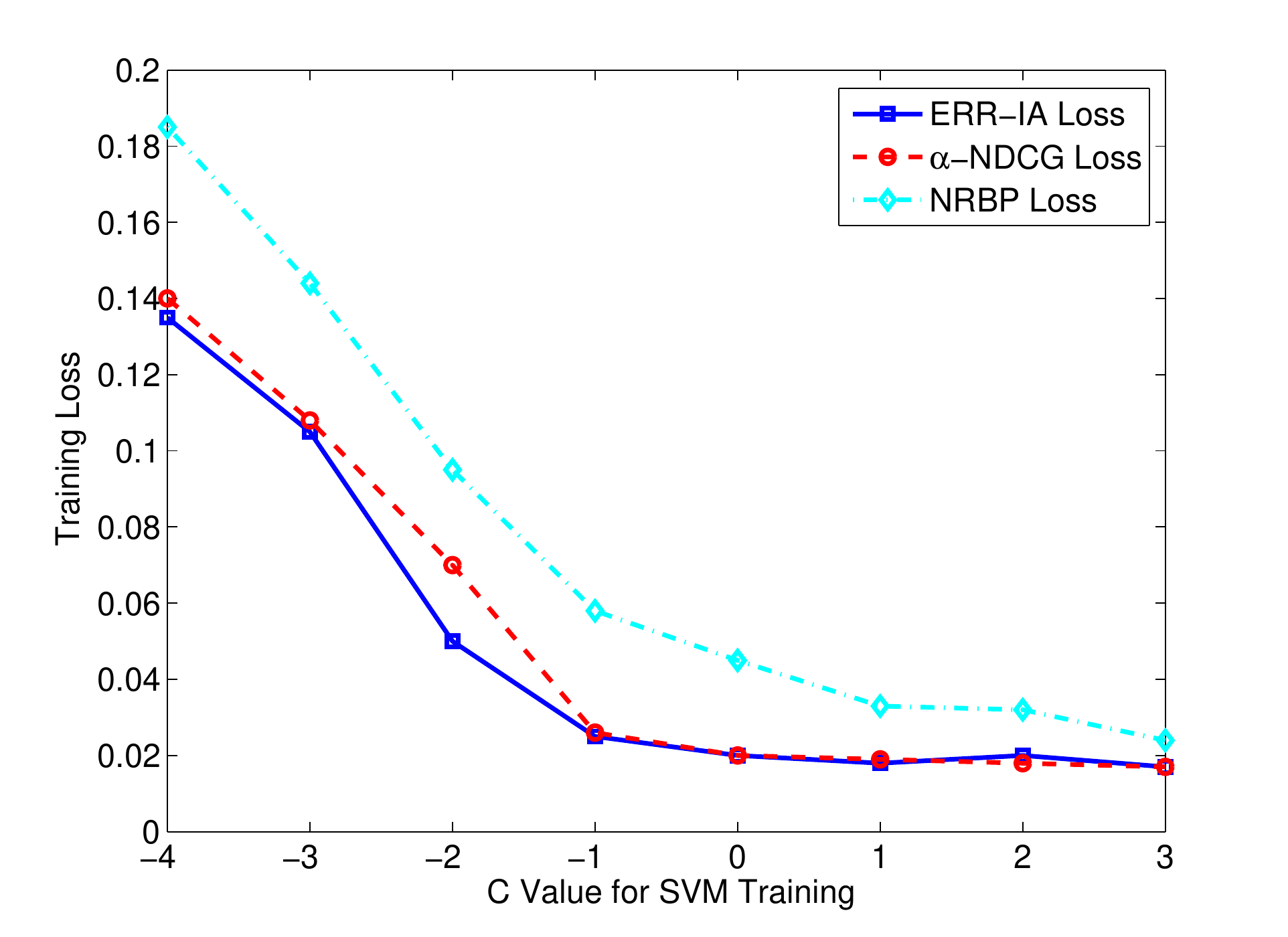}}
\caption{Training loss comparing $C$ values on WT2011 dataset.}
\label{fig:app}
\vspace{-0.2cm}
\end{figure}

We choose WT2011 as an example, and the training curves of our three models are shown in Figure~\ref{fig:app}. Obviously, with the increasing of $C$, all three models are able to fit the training data almost perfectly.  This indicates that our approximate constraint generation is acceptable for training purpose. The results for the other two datasets are similar, and we do not show them here due to space limitation.

\subsection{Feature Importance Analysis}

\begin{table*}[t]
\centering
\caption{Order list of diversity features with corresponding weight value. }
\label{tab:feature}
{
\begin{tabular}{c|ccccccc} \hline
feature & $\phi_{d_{odp}}$ & $\phi_{d_{topic}}$ & $\phi_{d_{title}}$ & $\phi_{d_{text}}$ & $\phi_{d_{anchor}}$ & $\phi_{d_{link}}$ & $\phi_{d_{url}}$ \\ \hline
weight & 2.82987 & 2.75189 & 0.95001& 0.87450 & 0.82735 & 0.06727 & 0.04800 \\
\hline
 \end{tabular}}
\vspace{-0.2cm}
\end{table*}

In this subsection, we will give some analysis on the importance of the proposed diversity features.
Table~\ref{tab:feature} shows the order list of features used in our learned model (SVM$_{\alpha\mbox{-}NDCG}$) according to the learned weight values (average on three datasets).
From the results, we can see that the $\phi_{d_{odp}}$ and $\phi_{d_{topic}}$ have been shown to be the most important, which is in accordance with our intuition that diversity mainly lies in the rich \textit{semantic information}.
Meanwhile, the title and anchor text diversity $\phi_{d_{title}}$ and $\phi_{d_{anchor}}$ also work well, since these fields typically provide a precise summary of the content of the document. Finally, The Link and URL based diversity $\phi_{d_{link}}$ and $\phi_{d_{url}}$ seem to be the least important features, which may be due to the sparsity of such types of features in the data.

As a learning-based method, our model is flexible to incorporate different types of features for capturing both the relevance and diversity. Therefore, it would be interesting to explore more other useful features to further improve performance of the diverse ranking. We will investigate the issue in future.

\subsection{Running Time Analysis}

We further study the efficiency of our approach and the baseline models.
All of the diversification methods (including the baseline models and our approach) associate with a greedy selection process,
which is time-consuming due to the consideration of the dependency relations of document pairs. Assuming that the size of output rankings is $K$, the size of candidate set is $n$, then this type of greedy selection based on maximizing a certain marginal gain, like Algorithm 1, will have time complexity of $O(n*K)$. With a small $K$, the running time is linear.

All the learning-based methods (i.e.~ListMLE, SVMDIV and SVM$_{DCEM}$) need additional offline training time due to the supervised learning process.
We compare the average training time of different learning-based methods, and the result is shown as Table~\ref{tab:time}:
\begin{table}
\caption{Average training time of different approaches. }
\label{tab:time}
{
\begin{tabular}{c|ccc} \hline
methods & ListMLE & SVMDIV & SVM$_{DCEM}$ \\ \hline
time (hours) & 1.5 & 2 & 2.5 \\
\hline
 \end{tabular}}
\end{table}

We can observe that our approach takes longer but comparable offline training time among different learning-based methods. Besides, in our experiments, we also found that the three variants of our SVM$_{DCEM}$ approach are with nearly the same training time. We will attempt to optimize our code to provide much faster training speed in the following work.


\section{Conclusions}

In this paper, we propose a unified structural learning framework for simultaneously optimizing both relevance and diversity. Firstly, we propose to directly use the diversity-correlated IR evaluation measures as the objective functions, such as ERR-IA, $\alpha$-NDCG and NRBP. Secondly, we define the discriminant function based on a bi-criteria objective to give consideration of both relevance and diversity. Thirdly, we propose and utilize a series of useful diversity-based features to facilitate the learning process.
Finally, we demonstrate empirically that our approach can significantly outperform the state-of-the-art methods on the public TREC datasets in all kinds of evaluation measures, and show better performance robustness.

Learning to optimize both relevance and diversity is an interesting direction. As for future work, we plan to take diversity into the consideration of the goal of traditional learning-to-rank framework. For example, we can add diversity-based score into the listwise loss functions \cite{Xia:20086}, to obtain a global ranked list which incorporates both relevance and diversity.

%
\bibliographystyle{abbrv}
\bibliography{sig-alternate}  

\begin{thebibliography}{10}

\bibitem{Agrawal:2009}
R.~Agrawal, S.~Gollapudi, A.~Halverson, and S.~Ieong.
\newblock Diversifying search results.
\newblock In {\em Proceedings of the Second ACM International Conference on Web
  Search and Data Mining}, WSDM '09, pages 5--14, 2009.

\bibitem{Brandt:2011:DRR}
C.~Brandt, T.~Joachims, Y.~Yue, and J.~Bank.
\newblock Dynamic ranked retrieval.
\newblock In {\em Proceedings of the fourth ACM international conference on Web
  search and data mining}, WSDM '11, pages 247--256, 2011.

\bibitem{MMR}
J.~Carbonell and J.~Goldstein.
\newblock The use of mmr, diversity-based reranking for reordering documents
  and producing summaries.
\newblock In {\em Proceedings of the 21st annual international ACM SIGIR
  conference on Research and development in information retrieval}, SIGIR '98,
  pages 335--336, 1998.

\bibitem{Carterette:2009:PMR}
B.~Carterette and P.~Chandar.
\newblock Probabilistic models of ranking novel documents for faceted topic
  retrieval.
\newblock In {\em Proceedings of the 18th ACM conference on Information and
  knowledge management}, CIKM '09, pages 1287--1296, 2009.

\bibitem{KDD08}
S.~Chakrabarti, R.~Khanna, U.~Sawant, and C.~Bhattacharyya.
\newblock Structured learning for non-smooth ranking losses.
\newblock In {\em Proceedings of the 14th ACM SIGKDD international conference
  on Knowledge discovery and data mining}, KDD '08, pages 88--96, 2008.

\bibitem{ERR}
O.~Chapelle, D.~Metlzer, Y.~Zhang, and P.~Grinspan.
\newblock Expected reciprocal rank for graded relevance.
\newblock In {\em Proceedings of the 18th ACM conference on Information and
  knowledge management}, CIKM '09, pages 621--630, 2009.

\bibitem{comparative}
C.~L. Clarke, N.~Craswell, I.~Soboroff, and A.~Ashkan.
\newblock A comparative analysis of cascade measures for novelty and diversity.
\newblock In {\em Proceedings of the fourth ACM international conference on Web
  search and data mining}, WSDM '11, pages 75--84, 2011.

\bibitem{TREC11}
C.~L. Clarke, N.~Craswell, I.~Soboroff, and E.~M.Voorhees.
\newblock Overview of the trec 2011 web track.
\newblock In {\em TREC}, 2011.

\bibitem{Clarke:2008}
C.~L. Clarke, M.~Kolla, G.~V. Cormack, O.~Vechtomova, A.~Ashkan,
  S.~B\"{u}ttcher, and I.~MacKinnon.
\newblock Novelty and diversity in information retrieval evaluation.
\newblock In {\em Proceedings of the 31st annual international ACM SIGIR
  conference on Research and development in information retrieval}, SIGIR '08,
  pages 659--666, 2008.

\bibitem{NRBP}
C.~L. Clarke, M.~Kolla, and O.~Vechtomova.
\newblock An effectiveness measure for ambiguous and underspecified queries.
\newblock In {\em Proceedings of the 2nd International Conference on Theory of
  Information Retrieval: Advances in Information Retrieval Theory}, ICTIR '09,
  pages 188--199, 2009.

\bibitem{proportionality}
V.~Dang and W.~B. Croft.
\newblock Diversity by proportionality: an election-based approach to search
  result diversification.
\newblock In {\em Proceedings of the 35th international ACM SIGIR conference on
  Research and development in information retrieval}, SIGIR '12, pages 65--74,
  2012.

\bibitem{Gollapudi:2009}
S.~Gollapudi and A.~Sharma.
\newblock An axiomatic approach for result diversification.
\newblock In {\em Proceedings of the 18th international conference on World
  wide web}, WWW '09, pages 381--390, 2009.

\bibitem{He:2012}
J.~He, V.~Hollink, and A.~de~Vries.
\newblock Combining implicit and explicit topic representations for result
  diversification.
\newblock In {\em Proceedings of the 35th international ACM SIGIR conference on
  Research and development in information retrieval}, SIGIR '12, pages
  851--860, 2012.

\bibitem{HeTMS12}
J.~He, H.~Tong, Q.~Mei, and B.~K. Szymanski.
\newblock Gender: A generic diversified ranking algorithm.
\newblock In {\em NIPS}, pages 1151--1159, 2012.

\bibitem{Hofmann:1999}
T.~Hofmann.
\newblock Probabilistic latent semantic indexing.
\newblock In {\em Proceedings of the 22nd annual international ACM SIGIR
  conference on Research and development in information retrieval}, SIGIR '99,
  pages 50--57, 1999.

\bibitem{rishabh2012-sub}
R.~Iyer and J.~Bilmes.
\newblock Submodular bregman divergences with applications.
\newblock In {\em Neural Information Processing Society (NIPS)}, Lake Taho, CA,
  December 2012.

\bibitem{Metzler:2005:MRF}
D.~Metzler and W.~B. Croft.
\newblock A markov random field model for term dependencies.
\newblock In {\em Proc. of the 28th annual international ACM SIGIR conference
  on Research and development in information retrieval}, SIGIR '06, pages
  472--479, 2005.

\bibitem{nemhauser1978analysis}
G.~Nemhauser, L.~Wolsey, and M.~Fisher.
\newblock An analysis of approximations for maximizing submodular set
  functions--i.
\newblock {\em Mathematical Programming}, 14(1):265--294, 1978.

\bibitem{qin:letor}
T.~Qin, T.-Y. Liu, J.~Xu, and H.~Li.
\newblock Letor: A benchmark collection for research on learning to rank for
  information retrieval.
\newblock {\em Inf. Retr.}, pages 346--374, 2010.

\bibitem{Radlinski:2006}
F.~Radlinski and S.~Dumais.
\newblock Improving personalized web search using result diversification.
\newblock In {\em Proceedings of the 29th annual international ACM SIGIR
  conference on Research and development in information retrieval}, SIGIR '06,
  2006.

\bibitem{Radlinski:2008:LDR}
F.~Radlinski, R.~Kleinberg, and T.~Joachims.
\newblock Learning diverse rankings with multi-armed bandits.
\newblock In {\em Proceedings of the 25th international conference on Machine
  learning}, ICML '08, pages 784--791, 2008.

\bibitem{Rafiei:2010}
D.~Rafiei, K.~Bharat, and A.~Shukla.
\newblock Diversifying web search results.
\newblock In {\em Proceedings of the 19th international conference on World
  wide web}, WWW '10, pages 781--790, 2010.

\bibitem{two-level}
K.~Raman, T.~Joachims, and P.~Shivaswamy.
\newblock Structured learning of two-level dynamic rankings.
\newblock In {\em Proceedings of the 20th ACM international conference on
  Information and knowledge management}, CIKM '11, pages 291--296, 2011.

\bibitem{Raman:2012:OLD2}
K.~Raman, P.~Shivaswamy, and T.~Joachims.
\newblock Online learning to diversify from implicit feedback.
\newblock In {\em Proceedings of the 18th ACM SIGKDD international conference
  on Knowledge discovery and data mining}, KDD '12, pages 705--713, 2012.

\bibitem{Sakai:2012:EIN}
T.~Sakai.
\newblock Evaluation with informational and navigational intents.
\newblock In {\em Proceedings of the 21st international conference on World
  Wide Web}, WWW '12, pages 499--508, 2012.

\bibitem{Sakai:2011}
T.~Sakai and R.~Song.
\newblock Evaluating diversified search results using per-intent graded
  relevance.
\newblock In {\em Proceedings of the 34th international ACM SIGIR conference on
  Research and development in Information Retrieval}, SIGIR '11, pages
  1043--1052, 2011.

\bibitem{Santos:ESR}
R.~L. Santos, P.~Jie, C.~Macdonald, and I.~Ounis.
\newblock Explicit search result diversification through sub-queries.
\newblock In {\em Proceedings of the 32nd European conference on Advances in
  Information Retrieval}, ECIR'2010, pages 87--99, 2010.

\bibitem{Santos:2010}
R.~L. Santos, C.~Macdonald, and I.~Ounis.
\newblock Exploiting query reformulations for web search result
  diversification.
\newblock In {\em Proceedings of the 19th international conference on World
  wide web}, WWW '10, pages 881--890, 2010.

\bibitem{Intent-aware}
R.~L. Santos, C.~Macdonald, and I.~Ounis.
\newblock Intent-aware search result diversification.
\newblock In {\em Proceedings of the 34th international ACM SIGIR conference on
  Research and development in Information Retrieval}, SIGIR '11, pages
  595--604, 2011.

\bibitem{icml/SlivkinsRG10}
A.~Slivkins, F.~Radlinski, and S.~Gollapudi.
\newblock Learning optimally diverse rankings over large document collections.
\newblock In {\em Proceedings of the 27th international conference on Machine
  learning}, ICML '10, pages 983--990, 2010.

\bibitem{Tsochantaridis:2005}
I.~Tsochantaridis, T.~Joachims, T.~Hofmann, and Y.~Altun.
\newblock Large margin methods for structured and interdependent output
  variables.
\newblock {\em J. Mach. Learn. Res.}, 6:1453--1484, Dec. 2005.

\bibitem{Explicit}
S.~Vargas, P.~Castells, and D.~Vallet.
\newblock Explicit relevance models in intent-oriented information retrieval
  diversification.
\newblock In {\em Proceedings of the 35th international ACM SIGIR conference on
  Research and development in information retrieval}, SIGIR '12, pages 75--84,
  2012.

\bibitem{Portfolio}
J.~Wang and J.~Zhu.
\newblock Portfolio theory of information retrieval.
\newblock In {\em Proceedings of the 32nd international ACM SIGIR conference on
  Research and development in information retrieval}, SIGIR '09, pages
  115--122, 2009.

\bibitem{Wang:2012:RRM}
L.~Wang, P.~N. Bennett, and K.~Collins-Thompson.
\newblock Robust ranking models via risk-sensitive optimization.
\newblock In {\em Proceedings of the 35th international ACM SIGIR conference on
  Research and development in information retrieval}, SIGIR '12, pages
  761--770, 2012.

\bibitem{Xia:20086}
F.~Xia, T.-Y. Liu, J.~Wang, W.~Zhang, and H.~Li.
\newblock Listwise approach to learning to rank: theory and algorithm.
\newblock In {\em Proceedings of the 25th international conference on Machine
  learning}, ICML '08, pages 1192--1199, 2008.

\bibitem{Yue:2008}
Y.~Yue and T.~Joachims.
\newblock Predicting diverse subsets using structural svms.
\newblock In {\em Proceedings of the 25th international conference on Machine
  learning}, ICML '08, pages 1224--1231, 2008.

\bibitem{Zhai:2003}
C.~X. Zhai, W.~W. Cohen, and J.~Lafferty.
\newblock Beyond independent relevance: methods and evaluation metrics for
  subtopic retrieval.
\newblock In {\em Proceedings of the 26th annual international ACM SIGIR
  conference on Research and development in informaion retrieval}, SIGIR '03,
  pages 10--17, 2003.

\bibitem{zhu:2014:rRLTR}
Y.~Zhu, Y.~Lan, J.~Guo, X.~Cheng, and N.~Shuzi.
\newblock Learning for search result diversification.
\newblock In {\em Proceedings of the 37th annual international ACM SIGIR
  conference on Research and development in information retrieval}, SIGIR'14,
  pages 293--302, 2014.

\end{thebibliography}

\end{document}